\documentclass[11pt,a4paper]{article}

\usepackage[no-natbib-sort]{my-jheppub}

\usepackage{graphicx}
\usepackage[english]{babel}
\usepackage{amsthm}

\usepackage[nodayofweek]{date time}

\def \ed  {\end{document} }

\begin{document}

\date{\currenttime}


\title{Chiral Higher Spin Theories and Self-Duality}

\emailAdd{d.ponomarev@imperial.ac.uk}

\author{Dmitry Ponomarev}

\affiliation{Theoretical physics group, Blackett Laboratory, Imperial College London,   SW7 2AZ, U.K.}


\abstract{We study recently proposed chiral higher spin theories  --- cubic theories of interacting
massless higher spin fields in four-dimensional flat space. We show that they are naturally associated 
with gauge algebras, which manifest themselves in several related ways. Firstly, 
the chiral higher spin equations of motion  can be reformulated as the self-dual Yang-Mills equations 
with the associated gauge algebras instead of the usual colour gauge algebra.
We also demonstrate that the chiral higher spin field equations, similarly to the self-dual Yang-Mills equations, feature an infinite
algebra of hidden symmetries, which ensures their integrability.
Secondly, we show that off-shell amplitudes in chiral higher spin theories satisfy the generalised BCJ relations with the
usual colour structure constants replaced by the structure constants of higher spin gauge algebras.
We also propose  generalised double copy procedures featuring higher spin theory amplitudes.
 Finally, using the light-cone deformation
procedure we prove that the structure of the Lagrangian that leads to all these properties is
universal and follows from Lorentz invariance.}

\unitlength = 1mm

\today
\date {}
\begin{flushright}\small{Imperial-TP-DP-2017-{01}}\end{flushright}

\maketitle

\def \lR {L}

\def \ll {{\cal \ell}}

\def \be {\begin{equation}}\def \ee {\end{equation}}

\def \ads {AdS$_4$\ }
\def \iffa {\iffalse} 

\def \ed {\bibliography{KHSA}{}
\bibliographystyle{utphys} \end{document}}

\section{Introduction}

There are numerous no-go results that rule out interactions of massless higher spin fields  in flat space
\cite{Weinberg:1964ew,Coleman:1967ad,Aragone:1979hx,Bekaert:2010hp,Dempster:2012vw,Joung:2013nma}\footnote{In three dimensions
due to absence of local degrees of freedom the standard no-go theorems do not apply and higher spin theories can be
constructed \cite{Afshar:2013vka,Gonzalez:2013oaa}.}. These results
rely on different approaches and different sets of assumptions,  see \cite{Bekaert:2010hw} for review. Recently,
this problem was reconsidered in \cite{Taronna:2017wbx,Roiban:2017iqg}.

Typically, one addresses the problem of interactions of higher spin gauge fields using the manifestly covariant
approach \cite{Berends:1984wp,Berends:1984rq}, that is the approach where Lorentz invariance is ensured by employment of Lorentz tensors.
Within the manifestly covariant approach  the only
non-trivial requirement one needs to achieve is perturbative gauge invariance of the action. The same problem can also be addressed
differently. In the light-cone deformation procedure one  deals exclusively with the dynamical degrees of freedom, while
Lorentz invariance is not manifest and should be imposed order by order in perturbations \cite{Bengtsson:1983pd,Bengtsson:1983pg,Bengtsson:1986kh}.
The reason that lead us to revisit the problem of interactions of massless higher spin fields in flat space is an observation that 
the light-cone deformation procedure allows to construct additional local cubic vertices compared to those available in the manifestly
covariant approaches. These additional vertices where known for some time \cite{Bengtsson:1986kh,Metsaev:1991mt,Metsaev:1991nb,Metsaev:1993ap},
but it is only recently that it was emphasised that they are missing in the Lorentz covariant classification 
\cite{Bengtsson:2014qza}\footnote{It is worth to note, that within the covariant approach there exist algebra deformations, which
cannot be promoted to the level of the action \cite{Boulanger:2006gr,Bekaert:2010hp}. Potentially, these may be related to the additional vertices of the light-cone approach.}.
At least formally, the additional vertices can be rewritten in the manifestly covariant form. However, this requires to introduce non-localities not only with
respect to space-time derivatives, but also with respect to operators that perform tensor contractions, see \cite{Conde:2016izb,Sleight:2016xqq}.

Similarly to other approaches, when proceeding  to higher orders in deformations, one finds that  consistency conditions that 
the light-cone deformation 
procedure imposes, become significantly harder to solve. Recently, some progress in this direction  was achieved when
it was established that the light-cone deformation procedure is closely related to the spinor-helicity approach. First, it was realised
\cite{Ananth:2012un,Akshay:2014qea} that upon proper rearrangement of total derivatives the light-cone cubic vertices can
 be brought to the form, which on-shell reproduces the known spinor-helicity classification \cite{Benincasa:2007xk}. This idea was
 extended to all orders in deformations in \cite{Ponomarev:2016cwi}, where it was shown that consistency of the light-cone deformation
 procedure is equivalent to the requirement that the off-shell amplitude built of the light-cone vertices in the standard way can be expressed in terms
 of spinor products of appropriate homogeneity degrees. This construction allows to derive an off-shell extension of the spinor-helicity approach,
 which for lower spin fields was introduced and employed previously, see \cite{Bardeen:1995gk,Cangemi:1996rx,Cachazo:2004kj,Gorsky:2005sf,Mansfield:2005yd,Brandhuber:2007vm,Ananth:2007zy}.

Having established equivalence between the light-cone deformation procedure and the spinor-helicity approach one can 
employ the no-go results for interactions of massless higher spin fields found within the latter framework
 \cite{Benincasa:2007xk,Benincasa:2011kn,Benincasa:2011pg,McGady:2013sga} to argue that such interactions
are also absent in the light-cone formalism. Independently, a local obstruction to the parity invariant minimal gravitational coupling of higher 
spin fields was found using the direct analysis of the light-cone consistency conditions \cite{PonSkv:2016}. 

In other words, there is a compelling  evidence that interacting local parity invariant higher spin theories do not exist
 even without an assumption of manifest Lorentz invariance. Of course, this conclusion was not totally unexpected, considering a long
 list of no-go theorems that we previously mentioned. In order to avoid these no-go results it was proposed 
 that higher spin fields can only interact in AdS space and some promising results were found \cite{Fradkin:1987ks,Fradkin:1986qy}.
 Additional evidence supporting this idea came from holography. In its simplest form the 
 conjecture made in \cite{Sezgin:2002rt,Klebanov:2002ja} relates higher spin theory in the bulk of AdS space with
  a free conformal theory on its boundary\footnote{We would like to point out that the idea that higher spin scattering
amplitudes in AdS can be extracted from correlators in the theory of the boundary conformal scalar
was put forward  \cite{Flato:1980zk} long before higher spin holography acquired its modern form.
Some explicit computations can be found in \cite{Fronsdal:1986ui}.}. This duality can be used rather efficiently to construct
the bulk higher spin theory order by order, as it was explicitly demonstrated in \cite{Bekaert:2014cea,Bekaert:2015tva,Sleight:2016dba}.
 Using this approach it was recently  found \cite{Sleight:2017pcz}  that the higher spin theory derived from holography features non-local 
quartic vertices. In a companion paper \cite{Ponomarev:2017qab} we elaborate further on these results and give a more quantitative description of non-locality
present in the holographic higher spin theory.
This analysis suggests that  the situation with locality of higher spin theories in AdS is
similar to that in flat space.

Though, non-locality of holographic higher spin theories may appear as a disappointing result,
 it is not yet clear whether it leads to any essential physical
pathologies. On the other hand, one can use this result in a positive way to learn more about locality and how it
 is violated in higher spin theories
in AdS, which,  in turn, may suggest a way to circumvent the locality problem  in flat space. We  believe that 
better understanding of the connection between higher spin theories in AdS and in flat space can also be useful for the 
theory on the AdS side, because flat space analysis does not have many of those technical  difficulties inherent to the analysis 
in AdS.

Despite the light-cone deformation procedure did not give any ways around locality obstructions for
parity invariant higher spin theories, it still lead to some interesting results. Based on the earlier analysis by 
Metsaev \cite{Metsaev:1991mt,Metsaev:1991nb,Metsaev:1991thesis} the chiral higher spin theory was
proposed in \cite{Ponomarev:2016lrm}. It is a cubic theory given by the action
\begin{align}
\notag
&S =\frac{1}{2} \sum_{\lambda_1,\lambda_2} \int d^4 q_1 d^4 q_2 \delta(q_1+q_2)\delta^{\lambda_1+\lambda_2,0} 
\Box_2\Phi^{\lambda_1}_1\Phi^{\lambda_2}_2\\
\label{27sep1}
&\;+\frac{g}{3}\sum_{\lambda_1,\lambda_2,\lambda_3}\int d^4 q_1 d^4 q_2 d^4 q_3 \delta(q_1+q_2+q_3)
\frac{\ell^{\lambda_1+\lambda_2+\lambda_3-1}}{(\lambda_1+\lambda_2+\lambda_3-1)!} \frac{\bar{\mathbb{P}}^{\lambda_1+\lambda_2+\lambda_3}_{23}}{\beta_1^{\lambda_1}\beta_2^{\lambda_2}\beta_3^{\lambda_3}}
\Phi_1^{\lambda_1}\Phi_2^{\lambda_2}\Phi_3^{\lambda_3},
\end{align}
where
\begin{equation}
\label{27sep2}
\Box_i \equiv 2(q_i^+ q_i^- + q_i\bar q_i), \qquad \bar{\mathbb{P}}_{ij} \equiv \bar q_i\beta_j -\bar q_j \beta_i , \qquad \beta_i \equiv q_i^+
\end{equation}
and $\lambda_i$ are helicities.
The chiral higher spin theory is consistent to all orders in deformations and satisfies a list
of attractive properties. First of all, it contains two-derivative couplings of higher spin fields to gravity, which can
be naturally regarded as minimal gravitational couplings.
 These vertices are absent in the manifestly covariant approach. Moreover,
it turns out that all higher spin fields couple to gravity with the same strength, which can be regarded as an extension
of Weinberg's equivalence principle \cite{Weinberg:1964ew}. This is despite the fact that the argument of Weinberg does
not literally apply here.

Yet another interesting feature of the chiral higher spin theory is that its coupling constants in some form agree
with those derived from holography.
 To be more precise, it was observed \cite{Skvortsov:2015pea} that 
for cubic vertices involving two scalars and a higher spin field there is a numerical agreement between coupling constants derived holographically 
 \cite{Bekaert:2015tva} and via the light-cone deformation procedure \cite{Metsaev:1991mt,Metsaev:1991nb,Metsaev:1991thesis}.
It was also conjectured that this pattern persists for more generic higher spin interactions. And, indeed, when the complete
cubic holographic action was found in \cite{Sleight:2016dba} the agreement with the light-cone formula for higher derivative
vertices was confirmed. This is the most that can be expected from such a comparison, because
 due to derivatives' ordering ambiguities, only higher derivative vertices in AdS have an unambiguous flat limit, see \cite{Boulanger:2008tg}.

Despite these remarkable properties, the chiral higher spin theory is far from being physically relevant. First of all, as its Hamiltonian is
not real, the associated evolution operator is not unitary. Moreover, in \cite{Ponomarev:2016lrm} it was found that the chiral
higher spin theory has a vanishing four-point function and is expected to have all higher point functions vanishing  as well. In this
respect it is similar to the self-dual Yang-Mills and the self-dual gravity theories\footnote{By the self-dual Yang-Mills/gravity
theory we will often mean not only the self-dual Yang-Mills/gravity equations, but also the associated actions. 
Lorentz invariance requires that these actions additionally feature fields of opposite helicities, see \cite{Chalmers:1996rq}. For
more details see sections \ref{sdym} and \ref{sect32}.}. 
In fact, the main result of this paper is that 
the chiral higher spin theory belongs to a class of theories all of which, similarly to self-dual gravity, can be interpreted as the generalised 
self-dual Yang-Mills theories. 
This connection with the self-dual Yang-Mills theory suggests that the chiral higher spin theory forms the self-dual sector
of some putative parity invariant higher spin  theory. Though, it is not yet clear how/if the latter can be constructed, already
now we may attempt to get some insights into its properties by studying its self-dual sector.
 For example, as in the case of Yang-Mills theory, one can expect that solutions of the self-dual theory also solve complete equations.

The main issue that we address in this paper is symmetries associated with chiral higher spin theories.
 Already in first papers on the covariant deformation procedure \cite{Berends:1984wp,Berends:1984rq}
 it was realised that deformations of the action should be accompanied with deformations of gauge transformations and of the algebra
 that they form. In the light-cone deformation procedure the relation between the action and  symmetries is less clear. 
 Contrary to the covariant approach, in 
 the light-cone approach gauge symmetry is absent for free theories and appearance of symmetries at 
 non-linear level, if any,  is due to a completely different mechanism. One possible way to proceed is
  to promote vertices to the covariant form and see what an algebra they induce
 \cite{Sleight:2016xqq}. This analysis leads to an interesting conclusion that the Metsaev couplings can be alternatively derived
 from the requirement that the structure constants of the algebra they induce coincide with the structure
 constants \cite{Joung:2014qya} of the Eastwood-Vasiliev algebra \cite{Eastwood:2002su,Vasiliev:2003ev}
  in the  Lorentz-like subsector.
 This approach, however, makes it necessary to deal with non-localities where they can be avoided. Moreover, it is not clear
 whether the resulting induced symmetry algebra characterises the initial light-cone theory or its particular covariant extension.
Finally, the analysis of  \cite{Sleight:2016xqq} was made in a parity invariant setup, whereas here 
our goal is to analyse higher spin theories and their symmetries in the self-dual sector.

In this paper we show that
the light-cone deformation procedure for massless fields, indeed, defines
the symmetry algebra, demonstrate the mechanisms how it appears
and clarify its meaning.
More precisely, we observe that simply  by dividing the  vertex of  the chiral higher spin theory by the kinematic part of the self-dual Yang-Mills
vertex we can construct structure constants that satisfy the Jacobi identity and, hence, define a Lie algebra. We call the algebra constructed via
this procedure  the {\em{gauge algebra}}.
For the chiral higher spin  theory it  is defined by the commutator
\begin{equation}
\label{27sep3}
\tilde E_1(x;z) =\sinh \left( \frac{\ell}{z} \left(\bar\partial_2 \partial^+_3- \partial^+_2  \bar\partial_3\right) \right)    \tilde E_2(x;z) \tilde E_3(x;z),
\end{equation}
where $\tilde E_i(x;z)$ are generating function for parameters of higher spin gauge transformations. These parameters depend
on all four space-time coordinates $x$ as well as on auxiliary variable $z$, introduced to combine gauge parameters associated
with different helicities into a single master gauge parameter. The former fact explains why we call this algebra a gauge one.

 The same algorithm allows to define gauge algebras
for a general class of theories, that naturally appears in the course of the analysis of the light-cone deformation procedure. All these theories
contain cubic vertices constructed only out of one type of spinors --- either angle or bracket ones --- and do not require completion by higher 
order vertices. Accordingly, we call them {\em{chiral cubic theories}}.

Next, we  show that by a procedure that can be regarded as a partial undoing of  light-cone gauge, the field equations for
this family of theories can be written as the self-duality conditions for the Yang-Mills curvature associated with a given gauge algebra.
 So, they can also be viewed as the
generalised self-dual Yang-Mills theories.
The self-duality condition for the Yang-Mills theory  leads to a list of special  properties such as an infinite hidden symmetry algebra,
integrability and an underlying twistor geometry, for review see \cite{Mason:1991rf,Hitchin:1999at}.
 We show that one of the constructions of the hidden symmetry algebra for the self-dual Yang-Mills and the self-dual gravity equations available in the
literature \cite{Park:1989vq,Park:1990fp} can be directly extended to include all chiral cubic theories. This also implies their integrability.

Another way how the structure of the Lagrangian of chiral cubic theories and the associated gauge algebras manifest
themselves is connected to the colour-kinematics duality 
\cite{Bern:2008qj,Bern:2010ue,Bern:2010yg}. This duality allows to organise perturbative computations in Yang-Mills 
theory and  connect them to gravity amplitudes via a simple
squaring procedure. An important element of this construction is the  Feynman-like
diagrammatic expansion of gauge theory amplitudes which involves only cubic vertices. 
 This expansion can be carried out in a way that the diagrams  it features satisfy a set
of relations, called the BCJ relations\footnote{It is 
 conjectured that the BCJ relations can be satisfied to all loop orders \cite{Bern:2008qj} and explicitly shown to the fourth loop order \cite{Bern:2013qca,Bern:2014sna}.
In this paper we are mainly concerned with tree diagrams.}.
  These relations, in turn, point towards the idea that along with the usual 
colour algebra,  Yang-Mills theory is controlled by the so-called kinematic algebra.
Moreover, once gauge theory amplitudes are rearranged into the form compatible with the BCJ relations,
the associated gravity amplitudes can be obtained by a simple double copy construction.

In the self-dual sector the colour-kinematics duality acquires a much simpler form, because the action from the
very beginning has only cubic vertices and the standard Feynman diagram expansion immediately produces amplitudes in 
the desired form. This makes structures relevant to the colour-kinematics duality manifest already at the level
of the action. Using this simplification, in \cite{Monteiro:2011pc} the kinematic algebra of the self-dual Yang-Mills theory
was identified as the  algebra of area-preserving diffeomorphisms. 
Since chiral cubic theories also have only cubic vertices, similar simplifications apply to them as well.
 In this paper we show that amplitudes in chiral cubic theories automatically satisfy relations that can be
regarded as the generalised BCJ relations. Moreover, we propose various  generalisations of the
squaring procedure that involve higher spin theory amplitudes.

This paper is organised as follows. In section \ref{caht} we review the light-cone deformation procedure. In particular, 
we emphasise that it implies the following property for massless theories in 4d flat space: once cubic holomorphic or antiholomorphic
vertices satisfy certain consistency conditions, then, by keeping only one set of  vertices and dropping the other, one 
obtains a complete theory. In other words, it is a general feature of the light-cone deformation procedure, that for chiral
theories it can be truncated at cubic order.
This property allows to define a class of chiral cubic theories.

Next, in section \ref{sect3} we propose a simple procedure that allows to extract  structure constants from given cubic vertices.
Then we consider numerous examples, which  illustrate how this procedure works and that the structure constants so defined,
 indeed, satisfy the Jacobi identity. 
In particular, in section \ref{section3.3} we discuss the chiral higher spin theory and the associated symmetry algebra.
Next, in section \ref{sect34} we give a similar analysis for the chiral higher spin theory with fields taking values in some internal
Lie algebra. In section \ref{sect3.5} we observe that the algebra associated with the chiral higher spin theory admits a contraction, similar
to the contraction that relates the differential operator algebra and the algebra generated by the Poisson bracket. 
By going backwards from structure constants to vertices
we find a chiral higher spin theory which, similarly to gravity, has only two-derivative interactions.

In the following two sections we interpret the structure of the Lagrangian and its connection to  Lie algebras that we found.
 First, in section \ref{Sect5} we show that the field equations for 
 chiral cubic  theories can be written in the form of self-duality conditions. These, in turn, can be rewritten as
 equations of motion for a certain sigma model. Hidden symmetries of this sigma model are well-known. This
 allows us to find the hidden symmetry algebras for chiral cubic  theories as well as to argue that they are integrable.
 Next, in section \ref{colkin}, we show that the pattern 
that we observed in previous sections has a natural interpretation in the context of colour-kinematics duality for self-dual theories.
We also suggest generalised double copy constructions involving chiral higher spin theories. 
 
 In section \ref{universality} we give a general argument that proves that the Jacobi identity for the gauge algebra
 structure constants is a consequence of the light-cone consistency conditions or, in other words, of Lorentz invariance.
 We finish the main part of the paper with concluding remarks in section \ref{conclusion}.
 
 This paper also has three appendices. In Appendix \ref{conventions} we collect our notations.
  In Appendix \ref{gsp} we show how off-shell self-dual Yang-Mills amplitudes can be related to those
 of chiral higher spin theories. Finally,  
 in Appendix \ref{locobstr} we review a local obstruction to the  minimal  parity invariant  gravitational coupling of
  higher spin fields found in \cite{PonSkv:2016}.

\section{Chiral cubic theories}
\label{caht}
In \cite{Metsaev:1991mt} it was realised that
for massless  theories in 4d flat space formulated in  light-cone gauge,
 Lorentz invariance implies two relatively simple constraints featuring only the cubic 
 action: one constraint for the holomorphic part of the action and one for the
 antiholomorphic part. This argument was simplified in  \cite{Ponomarev:2016lrm},
 as well as it was pointed out that keeping only one part of the cubic action
 --- either holomorphic or antiholomorphic --- allows to 
 obtain a consistent theory, which does not require any completion with
 higher order vertices.  In this way one finds a simple family of theories, which we
will call {\em{chiral cubic theories}}. The chiral higher spin theory proposed in
\cite{Ponomarev:2016lrm} is a particular example of a theory from this class. 
In the remainder
of this section we make a brief review of relevant results in \cite{Bengtsson:1983pd,Bengtsson:1983pg,Bengtsson:1986kh,Metsaev:1991mt,Metsaev:1991nb,Ponomarev:2016lrm}.
For a more comprehensive review of basics of the light-cone deformation procedure
in notations consistent with those used here, see  \cite{Ponomarev:2016cwi}.

 \subsection{Free theory}
 \label{sect2.1}
In  light-cone gauge the free action for a set of massless fields of helicities $\lambda$ is given by
 \begin{equation}
\label{1aug1}
S_2\equiv \int d^4 x L_2(x), \qquad L_2(x) = -\frac{1}{2}\sum_{\{\lambda,\alpha\}}  \partial_a\Phi^{-\lambda|\alpha}(x) \partial^a \Phi^{\lambda|\alpha}(x),
\end{equation}
where $\alpha$ enumerates internal degrees of freedom and we do not impose any constraints on the spectrum except that 
 opposite helicities are required to enter in pairs. Here and below we assume that the invariant norm in internal
 space is given by a unit metric $\delta_{\alpha\alpha}$. As usual, we will not distinguish vectors and 1-forms in internal space: they will both
 carry upper indices. For other conventions, see Appendix \ref{conventions}.

 As can be  seen from (\ref{1aug1}), all fields of non-zero helicities
 have identical Lagrangians. The only difference between them is how they
 transform with respect to the Poincare algebra, namely,
\begin{align}
\notag
P^i \cdot \Phi^{\lambda|\alpha}&\,\equiv \partial^i \Phi^{\lambda|\alpha},\\
\label{1aug2}
 J^{ij}\cdot \Phi^{\lambda|\alpha}&\,\equiv (x^i\partial^j - x^j \partial^i+S^{ij} )\Phi^{\lambda|\alpha}.
\end{align}
Here $S^{ij}$ is the spin part of the angular momentum operator, which can be defined from 
\begin{equation}
\label{1aug3}
S^{+i}\cdot \Phi^{\lambda|\alpha} =0,\qquad  S^{ij} \partial_i \cdot \Phi^{\lambda|\alpha}=0
\end{equation}
and by requiring  that the fields transform properly under the Wigner little group rotations 
\begin{equation}
\label{1aug4}
S^{x\bar x} \cdot \Phi^{\lambda|\alpha} = -\lambda \Phi^{\lambda|\alpha}.
\end{equation}
It is not hard to check that (\ref{1aug2}) are, indeed, the symmetries of the acton (\ref{1aug1}).

These symmetries are associated with the canonical  Noether currents 
 \begin{align}
\notag
 P^i \quad \to  \;\quad T^{i,j} &\,= \sum_{\{\lambda,\alpha\}}\frac{\delta L_2}{\delta (\partial_j \Phi^{\lambda|\alpha})}\partial^i \Phi^{\lambda|\alpha} - \eta^{ij} L_2,\\
 J^{ij} \quad   \to  \quad L^{ij,k} &\,= x^i T^{j,k}-x^j T^{i,k}+R^{ij,k},
  \label{1aug5}
 \end{align}
where $R^{ij,k}$ is the spin current
  \begin{equation}
  \label{1aug6}
  R^{ij,k} = \sum_{\{\lambda,\alpha\}}\frac{\delta L_2}{\delta( \partial_k \Phi^{\lambda|\alpha})} S^{ij}\cdot \Phi^{\lambda|\alpha}
  \end{equation}
 and  $S^{ij}$ was defined in (\ref{1aug3}), (\ref{1aug4}). Integrating them over a constant light-cone
 time surface $x^+=0$ we obtain the Noether charges
  \begin{equation}
  \label{1aug7}
  P^i_2 = \int d^3x^\perp T^{i,+}, \qquad J^{ij}_2 = \int d^3x^\perp L^{ij,+},
  \end{equation}
 which with a slight abuse of terminology we denoted by the same symbols as the associated 
 symmetry generators. These charges generate the Poincare algebra with respect to the
 Dirac bracket given by 
  \begin{equation}
  \label{1aug8}
 [\Phi^{\lambda|\alpha}(x^\perp,x^+),\Phi^{\mu|\beta}(y^\perp,x^+)]= \frac{1}{\partial^+_x - \partial^+_y}\delta^{\alpha\beta}\delta^{\lambda+\mu,0}\delta^{3}(x^\perp,y^\perp).
 \end{equation}
 
 In the following we will often use the Fourier transform with respect to spatial coordinates $x^-$, $x$ and $\bar x$
  \begin{align}
\notag
  \Psi(x,x^+) &\,= (2\pi)^{-\frac{3}{2}} \int{e^{+i (x^- p^+ + \bar x p +x\bar p)}\Psi(p,x^+)d^{3}p^\perp},\\
    \label{1aug9}
   \Psi(p,x^+) &\,= (2\pi)^{-\frac{3}{2}} \int{e^{-i (x^- p^+ + \bar x p +x\bar p)}\Psi(x,x^+)d^{3}x^\perp},
  \end{align}
 followed by a change of variables $p=iq$. Effectively, this amounts to the replacement
  \begin{equation}
  \label{1aug10}
  \frac{\partial}{\partial x^i} \to q_i, \qquad x^i \to -\frac{\partial}{\partial q_i}.
  \end{equation}
 We hope that our notation in (\ref{1aug9}) that employs the same symbol for a function and its Fourier 
 transform will not result in confusions.

 \subsection{Interactions}
 At the interacting level the action (\ref{1aug1}) receives higher order corrections. Consistency requires that these
 do not break Poincare symmetry of the theory. In other words, in order to ensure consistency, one has to deform
 the charges of the free theory (\ref{1aug7})  with non-linear terms and make sure that they still generate the 
 Poincare algebra. Implementation of this condition order by order is the main idea of the light-cone deformation
 procedure  \cite{Bengtsson:1983pd,Bengtsson:1983pg}.
 
 It is well known that it suffices to deform only the generators, that are tangent to the light-cone $x^+=0$ \cite{Dirac:1949cp}.
 These are 
 \begin{equation}
 \label{1aug11}
H\equiv P^- ,\qquad J\equiv J^{x-}, \qquad \bar J\equiv J^{\bar x-} 
\end{equation}
 and they are called \emph{dynamical}. The remaining generators are called \emph{kinematical}.
 
 Explicitly the dynamical generators at the free level are given by
 \begin{align}
\notag
H_2 &\,= \sum_{\{\lambda,\alpha\}}\int d^3q^\perp_1 d^3q^\perp_2 \delta^3(q^\perp_1+q^\perp_2)\beta_1 \Phi^{-\lambda|\alpha}(q^\perp_1,x^+) h_2^{\lambda|\alpha}(q^\perp_2) \Phi^{\lambda|\alpha}(q^\perp_2,x^+),\\
\label{1aug12}
J_2 &\,= \sum_{\{\lambda,\alpha\}}\int d^3q^\perp_1 d^3q^\perp_2 \delta^3(q^\perp_1+q^\perp_2)\beta_1 \Phi^{-\lambda|\alpha}(q^\perp_1,x^+) j_2^{\lambda|\alpha}(q^\perp_2) \Phi^{\lambda|\alpha}(q^\perp_2,x^+),
\end{align}
 where
 \begin{equation}
 \label{1aug13}
 h_2^{\lambda|\alpha}(q^\perp_2) = -\frac{q\bar q}{\beta}, \qquad j_2^{\lambda|\alpha}(q^\perp_2) = \frac{\partial}{\partial \bar q}\frac{q\bar q}{\beta}+ q \frac{\partial}{\partial \beta}+ \lambda \frac{q}{\beta}.
 \end{equation}
The charge $\bar J$ at every order is just a complex conjugate of $J$ and will not be specified  separately.

At the non-linear level one has 
\begin{equation}
\label{1aug14}
H=H_2+\sum_n H_n ,\quad J=J_2+\sum_n J_n
\end{equation}
with
 \begin{align}
 \notag
  H_n  =&\;  \frac{g^{n-2}}{n!}\sum_{\{\lambda_i,\alpha_i\}}\int d^{3n}q^\perp \delta^3 (\sum_{i=1}^n q^\perp_i) h^{\lambda_1|\alpha_1, \dots ,\lambda_n|\alpha_n}_n(q^\perp_i) \prod_{i=1}^n\Phi^{\lambda_i|\alpha_i}(q^\perp_i, x_i^+),\\
    \notag
  J_n  =&\;   \frac{g^{n-2}}{n!}\sum_{\{\lambda_i,\alpha_i\}}\int d^{3n}q^\perp \delta^3 (\sum_{i=1}^n q^\perp_i) \Big[j_n^{\lambda_1|\alpha_1, \dots ,\lambda_n|\alpha_n}(q^\perp_i)  
  \\  \label{1aug15}
  &\qquad\qquad\qquad\qquad\qquad   -\frac{1}{n}
  h_n^{\lambda_1|\alpha_1,\dots ,\lambda_n|\alpha_n}(q^\perp_i) \big( \sum_j \frac{\partial}{\partial \bar q_j}\big)\Big]\prod_{i=1}^n\Phi^{\lambda_i|\alpha_i}(q^\perp_i,x_i^+),
\end{align}
where $g$ is the coupling constant. The ansatz (\ref{1aug15}) is just the most general one that takes into account the following two facts.
First, on-shell $q^-$ dependence can always be eliminated employing free equations of motions. Similarly, off-shell it can be eliminated
by virtue of field redefinitions. Second, translation invariance requires $H_n$ to be proportional to the momentum conserving 
delta-function and $h_n$  be independent of space-time coordinates. In the Fourier representation the latter condition
translates into  a requirement that $h_n$ is free of derivatives with respect to momentum. Analogous considerations fix the form of the ansatz for $J_n$.
Let us also note that given our choice of $h_n$ to be $q^-$ independent, the deformation of the action is simply related to the
deformation of the Hamiltonian as follows
\begin{equation}
\label{1aug015}
\delta S_n = -\int dx^+ H_n.
\end{equation}

Next one ensures that dynamical generators commute with kinematical ones as required by the Poincare algebra.
This imposes the following constraints on $h_n$
\begin{align}
\notag
\delta^3 (\sum_{i=1}^n q^\perp_i)\sum_{i=1}^n \beta_i \frac{\partial}{\partial \bar q_i} h_n^{\lambda_1,\dots ,\lambda_n}&=0,
\quad \;
\delta^3 (\sum_{i=1}^n q^\perp_i)\sum_{i=1}^n \beta_i \frac{\partial}{\partial  q_i} h_n^{\lambda_1,\dots ,\lambda_n}&=0,\\
\delta^3 (\sum_{i=1}^n q^\perp_i)\sum_{i=1}^n (N_{q_i}- N_{\bar{q}_i}+\lambda_i)h_n^{\lambda_1,\dots, \lambda_n}&=0,
\quad \;
\delta^3 (\sum_{i=1}^n q^\perp_i) \sum_{i=1}^n \beta_i \frac{\partial}{\partial \beta_i} h_n^{\lambda_1,\dots ,\lambda_n}&=0,
 \label{1aug16}
\end{align}
where
\begin{equation}
\label{25sep1}
N_{q_i}\equiv q_i \frac{\partial}{\partial q_i}, \qquad  N_{\bar q_i}\equiv \bar q_i \frac{\partial}{\partial \bar q_i}
\end{equation}
and for brevity we omitted $q$ and $\alpha$ dependence of $h_n$. It will be reinstated when necessary.
Constrains on $j_n$ are similar to (\ref{1aug16})
 and will not be specified. They can be found, for example, in \cite{Ponomarev:2016lrm}.
 
 The first line in (\ref{1aug16}) implies that $h_n$ can contain momenta $\bar q_i$ and $q_i$
only in the following combinations with $\beta_j$
\begin{equation}
\label{1aug17}
\bar{\mathbb{P}}_{ij} \equiv \bar q_i\beta_j -\bar q_j \beta_i , \qquad \mathbb{P}_{ij} \equiv q_i\beta_j -q_j \beta_i.
\end{equation}
The remaining constraints relate homogeneity degrees of momenta with helicities of the fields entering the charge.

Less trivial is to ensure that dynamical generators commute appropriately. We will focus on the commutator
\begin{equation}
  \label{1aug18}
 [H,J]=0 \qquad \Leftrightarrow \qquad [H_2, J_n] + [H_3, J_{n-1}] + \dots + [H_{n-1}, J_3]+ [H_n,J_2]=0.
  \end{equation}
The commutator $[H,\bar J]=0$ is analogous and $[J,\bar J]=0$ is fulfilled automatically as a consequence
of the first two  \cite{Ponomarev:2016lrm}. We can readily compute
\begin{align}
  [H_2, J_n]  
  =  -\frac{g^{n-2}}{n!}\sum_{\lambda_i}\int d^{3n}q^\perp \delta (\sum_{i=1}^n q^\perp_i)  
  {\cal H}^{\{i\}}
 \Big[j_n^{\lambda_1, \dots ,\lambda_n} +\frac{1}{n}
 \big( \sum_j \frac{\partial}{\partial \bar q_j}\big) h_n^{\lambda_1,\dots, \lambda_n}\Big]
 \prod_{i=1}^n\Phi^{\lambda_i}(q_i),
  \label{1aug19}
  \end{align}
and 
\begin{align}
  \label{1aug20}
 [H_n,J_2] = \frac{g^{n-2}}{n!}\sum_{\lambda_i}\int d^{3n}q^\perp \delta (\sum_{i=1}^n q^\perp_i){\cal J}^{\{i\}} h_n^{\lambda_1, \dots ,\lambda_n} 
 \prod_{i=1}^n\Phi^{\lambda_i}(q_i),
 \end{align}
 where 
  \begin{equation}
  \label{1aug21}
  {\cal H}^{\{i\}}\equiv \sum_{i=1}^n h_2^{\lambda_i|\alpha_i}(q^\perp_i), \qquad {\cal J}^{\{i\}} =\sum_{i=1}^n \Big(-\frac{q_i\bar q_i}{\beta_i}\frac{\partial}{\partial \bar q_i}-q_i \frac{\partial}{\partial \beta_i}+\lambda_i \frac{q_i}{\beta_i}\Big).
  \end{equation}
 
\subsection{Cubic vertices}
At the first non-trivial order we need to solve
\begin{equation}
\label{1aug22}
[H_2,J_3]+[H_3,J_2]=0.
\end{equation}
One way to do this is to note \cite{Ponomarev:2016cwi} that by using momentum conservation inside $h_n$ 
one can always achieve 
\begin{equation}
\label{1aug23}
j_n^{\lambda_1, \dots ,\lambda_n} +\frac{1}{n}
 \big( \sum_j \frac{\partial}{\partial \bar q_j}\big) h_n^{\lambda_1,\dots, \lambda_n}=0.
\end{equation}
This implies the following two consequences. Firstly,
\begin{equation}
\label{1aug24}
[H_2,J_n]=0.
\end{equation}
Secondly, it determines $j_n$ in terms of $h_n$ to all orders.

With (\ref{1aug23}) imposed,  (\ref{1aug22}) reduces to 
\begin{equation}
\label{1aug25}
\delta^3 (\sum_{i=1}^n q^\perp_i) \sum_{i=1}^n {\cal J}^{\{i\}}h_3^{\lambda_1,\lambda_2,\lambda_3}=0,
\end{equation}
where ${\cal J}$ was defined in (\ref{1aug21}).
This condition should be supplemented with an analogous one
\begin{equation}
\label{1aug025}
\delta^3 (\sum_{i=1}^n q^\perp_i) \sum_{i=1}^n \bar{\cal J}^{\{i\}}h_3^{\lambda_1,\lambda_2,\lambda_3}=0,
\end{equation}
 that follows from the commutator $[H_3,\bar J_2]=0$.
 
There is a simple way to solve (\ref{1aug25}), (\ref{1aug025}) simultaneously, which naturally leads to the spinor-helicity representation and
directly extends to all orders \cite{Ponomarev:2016cwi}. 
 For our current purposes, however,  the spinor-helicity representation will not be needed and
it is more economical to present the cubic vertices in the form, see \cite{Metsaev:1991mt}
\begin{equation}
\label{2aug3}
h^{\lambda_1,\lambda_2,\lambda_3}_3 = -{C^{\lambda_1,\lambda_2,\lambda_3}}
\frac{\bar{\mathbb{P}}^{\lambda_1+\lambda_2+\lambda_3}}{\beta_1^{\lambda_1}\beta_2^{\lambda_2}\beta_3^{\lambda_3}}
-
{\tilde C^{-\lambda_1,-\lambda_2,-\lambda_3}} \frac{\mathbb{P}^{-\lambda_1-\lambda_2-\lambda_3}}{\beta_1^{-\lambda_1}\beta_2^{-\lambda_2}\beta_3^{-\lambda_3}},
\end{equation}
where
\begin{equation}
\label{2aug2}
\bar{\mathbb{P}}\equiv \bar{\mathbb{P}}_{12}=\bar{\mathbb{P}}_{23}=\bar{\mathbb{P}}_{31}, \qquad \mathbb{P}\equiv \mathbb{P}_{12}=\mathbb{P}_{23}=\mathbb{P}_{31}
\end{equation}
and $C$, $\tilde C$ control dependence of the coupling constants on helicity and internal labels.
In the following we will call the first term in (\ref{2aug3}) \emph{antiholomorphic},
while the second term will be called \emph{holomorphic}. 
To determine $j_3$ from $h_3$ one can use (\ref{1aug23}). This will eventually specify all deformed charges and the 
deformed action (\ref{1aug015}). Note, however, that to determine $j_3$ from (\ref{1aug23}), first, one has to rewrite (\ref{2aug3})
 in the spinor-helicity form. How to do that  is briefly reviewed in Appendix \ref{conventions}.

Let us make a couple of comments about the structure of light-cone cubic vertices, which will be useful in the following.
Firstly, as it is not hard to see, for the case of trivial colour dependence, the cubic Hamiltonian 
density $h$  (\ref{2aug3}) is symmetric/antisymmetric with respect to permutations of field labels if
the total helicity in the vertex is even/odd. At the same time, the fields themselves are bosons, so when the hamiltonian density is integrated against them it gets projected into its totally symmetric part.
 This implies that in the case of trivial colour dependence vertices with the total helicity being odd effectively drop out.
One can also consider the opposite case --- when fields take values in some Lie algebra and vertices are proportional
to the totally antisymmetric Lie algebra structure constants. Then, contrary to the uncoloured case, the vertices with
 the total helicity being even drop out.

Also, note that (\ref{1aug25}), (\ref{1aug025})  admit solutions that involve 
terms of the form $\bar{\mathbb{P}}\mathbb{P}$. These were dropped, because they vanish when evaluated on solutions
of free equations of motion.
Indeed, one can see that
\begin{equation}
\label{2aug4}
\bar{\mathbb{P}}\mathbb{P} = -\beta_1\beta_2\beta_3 \sum_{i=1}^3 \frac{q_i\bar q_i}{\beta_i}\approx 
\beta_1\beta_2\beta_3 \sum_{i=1}^3 q_i^-=0,
\end{equation}
 where "$\approx$" denotes  equalities that hold up to terms, proportional to free equations of motion.

Finally,  in the light-cone analysis one typically imposes locality by requiring that interactions are polynomial in $q$ and $\bar q$. 
Even without detailed analysis --- just
on dimensional grounds --- one can easily see that the total power of $q$ and $\bar q$ in the light-cone vertex equals to the number of derivatives in its covariant form, provided the latter form exists. Thus, the light-cone locality defined above can be regarded as a natural counterpart of locality in the covariant formulation, defined as the requirement that interactions have finitely many derivatives. Let us, however, remind the reader that some of the light-cone vertices cannot be promoted to the covariant form and remain local, which means that these two definitions of locality  are not equivalent.

With this clarified, we impose light-cone locality, which implies
\begin{align}
\notag
C^{\lambda_1,\lambda_2,\lambda_3}&=0, \qquad \text{for} \qquad \lambda_1+\lambda_2+\lambda_3< 0,\\
\label{2aug5}
\tilde C^{\lambda_1,\lambda_2,\lambda_3}&=0, \qquad \text{for} \qquad \lambda_1+\lambda_2+\lambda_3 > 0.
\end{align}
The case $\lambda_1+\lambda_2+\lambda_3=0$ is somewhat special and is treated differently in 
different approaches. In the spinor-helicity approach when the total helicity is zero both holomorphic and antiholomorphic
vertices are allowed \cite{Benincasa:2007xk}. In contrast, in the light-cone deformation procedure one can check that $j_3$ and $\bar j_3$
are non-local unless $\lambda_1=\lambda_2=\lambda_3=0$, see e.g.  \cite{Metsaev:1991mt}.
 In other words, from the light-cone analysis perspective, 
only a scalar self-interaction $\Phi^3$ is a local interaction with the total helicity zero.

In addition, to obtain a real Hamiltonian one should demand
\begin{equation}
\label{2aug6}
\tilde C^{\lambda_1,\lambda_2,\lambda_3}= \bar C^{-\lambda_1,-\lambda_2,-\lambda_3},
\end{equation}
which results into a parity invariant theory.
Otherwise, coupling constants at this order of perturbation theory are arbitrary.

\subsection{Truncation at cubic order}
\label{sect24}
At the next order in deformation we need to solve
\begin{equation}
\label{2aug7}
[H_3,J_3]+[H_4,J_2]=0,
\end{equation}
where we used (\ref{1aug24}) to drop $[H_2,J_4]$. Slightly rephrasing an observation made in \cite{Metsaev:1991mt},
we note that $[H_4,J_2]$, if non-vanishing, is at least linear in $q$, see (\ref{1aug21}). Then, (\ref{2aug7})
 implies that the $q$-independent part of $[H_3,J_3]$ should vanish separately
\begin{equation}
\label{2aug8}
[H_3,J_3]\Big |_{q=0}=0.
\end{equation}
It is not hard to see that only the antiholomorphic part of the vertex (\ref{2aug3}) contributes to (\ref{2aug8}),
which, thus, results in a quadratic equation for coupling constants $C^{\lambda_1,\lambda_2,\lambda_3}$.

Analogously, focusing on the conjugate consistency condition
\begin{equation}
\label{2aug9}
[H_3,\bar J_3]+[H_4,\bar J_2]=0 
\end{equation}
and repeating the same arguments we arrive at
\begin{equation}
\label{2aug10}
[H_3,\bar J_3]\Big |_{\bar q=0}=0,
\end{equation}
which results in a quadratic constraint for $\tilde C^{\lambda_1,\lambda_2,\lambda_3}$.

Generically, both $C$ and $\tilde C$ are simultaneously non-zero. This certainly happens if we demand
that the Hamiltonian is real and the theory is parity invariant, see (\ref{2aug6}). In this case (\ref{2aug8}) does not imply
 $[H_3,J_3]=0$, so (\ref{2aug7}) requires a non-zero quartic Hamiltonian $H_4$. Similarly,
 generically, (\ref{2aug9}) does not imply $[H_3,\bar J_3]=0$ and (\ref{2aug9}) gives
 another equation for $H_4$ to satisfy.
 
 However, as was noted in \cite{Ponomarev:2016lrm}, if we take, say, 
 \begin{equation}
 \label{2aug11}
 \tilde C^{\lambda_1,\lambda_2,\lambda_3}=0,
 \end{equation} 
 then (\ref{2aug8}) implies $[H_3,J_3]=0$ and (\ref{2aug7}) can be solved by setting the 
 quartic vertex to zero
 \begin{equation}
 \label{2aug12}
 h_4=0.
 \end{equation}
On the other hand, for (\ref{2aug11}) one also has $[H_3,\bar J_3]=0$, so (\ref{2aug11}), (\ref{2aug12}) also solve (\ref{2aug9}).
Clearly, by setting all higher vertices to zero consistency conditions at higher orders  are trivially satisfied and
we end up with a consistent theory featuring only cubic antiholomorphic vertices\footnote{Analogous phenomenon of 
truncation of the deformation procedure was noticed using the approach of unfolding  for self-dual gravity in AdS \cite{Vasiliev:1989xz}.
To the best of our knowledge this analysis has not been extended to higher spin theories.}. Similarly, one can construct a theory
featuring only cubic holomorphic vertices.

Let us summarise the discussion of this section. We considered the light-cone deformation procedure
as a general framework to construct interactions of massless fields in 4d flat space. At first non-trivial order
one introduces cubic vertices, which  split into the holomorphic and the antiholomorphic parts.
Consistency of the light-cone deformation procedure implies 
 that, irrespectively of higher order terms,  antiholomorphic vertices
should satisfy (\ref{2aug8}), while  holomorphic ones should satisfy (\ref{2aug10}). 
Once these conditions are satisfied, each set of vertices separately --- holomorphic or antiholomorphic ones ---
defines a cubic theory, which is consistent to all orders in the coupling constant.
Explicitly, the complete action in the antiholomorphic case is of the form
\begin{align}
\notag
&S =\frac{1}{2} \sum_{\lambda_1,\lambda_2} \int d^4 q_1 d^4 q_2 \delta(q_1+q_2)\delta^{\lambda_1+\lambda_2,0} 
\Box_2\Phi^{\lambda_1}_1\Phi^{\lambda_2}_2\\
\label{6aug1}
&\qquad \quad +\frac{g}{3!}\sum_{\lambda_1,\lambda_2,\lambda_3}\int d^4 q_1 d^4 q_2 d^4 q_3 \delta(q_1+q_2+q_3)
C^{\lambda_1,\lambda_2,\lambda_3} \frac{\bar{\mathbb{P}}^{\lambda_1+\lambda_2+\lambda_3}_{23}}{\beta_1^{\lambda_1}\beta_2^{\lambda_2}\beta_3^{\lambda_3}}
\Phi_1^{\lambda_1}\Phi_2^{\lambda_2}\Phi_3^{\lambda_3},
\end{align}
where $\Box_i$ was defined in (\ref{27sep2}).
Cubic holomorphic and  antiholomorphic theories will be jointly called chiral cubic theories.

A paradigmatic example of a chiral cubic theory is given by the self-dual Yang-Mills
theory with the action proposed by Chalmers and Siegel \cite{Chalmers:1996rq}. It describes the self-dual Yang-Mills field as well as the
 field of opposite helicity propagating on the self-dual background. 
 In the following section we will demonstrate that all chiral cubic theories and the self-dual Yang-Mills theory in particular have 
 a very special form of $C^{\lambda_1,\lambda_2,\lambda_3}$ that allows to associate these theories with Lie
 algebras.

\section{Structure constants from cubic vertices}
\label{sect3}
 In this section we illustrate that chiral cubic  theories  are associated with Lie algebras in a very simple way.
 This will provide a first step towards identification of them as  generalised self-dual Yang-Mills theories.
We begin by giving a simple prescription of how to define Lie algebra structure constants from
vertices of chiral cubic  theories. 
 Then for a representative set of examples\footnote{A complete and fully 
 rigorous classification of chiral  cubic theories has not yet been done, though,
 a systematic way to solve the associated consistency conditions was given in \cite{Ponomarev:2016lrm}.
 It suggests that the list of theories that we consider below is, in essence, an exhaustive one. What it misses is 
only theories that can be obtained from theories that we consider by contractions or truncations of the spectrum of fields.} we demonstrate  that the structure constants so defined  do satisfy
 the Jacobi identity.
 A more universal argument proving that the Jacobi identity is satisfied as a consequence of the light-cone consistency conditions
  will be given in section \ref{universality}.

Roughly speaking, for a given cubic chiral theory structure constants of the associated gauge
algebra will be defined as a ratio of its cubic vertex  and the vertex of the
self-dual Yang-Mills theory. Below this definition will be given in a bit more formal way. 
In this definition we would like to emphasise that structure constants are more naturally connected
to the equations of motions rather than to the action these equations are derived from.
The reason is very simple: a Lie algebra defines a bilinear product, while a cubic vertex defines a three-form.
To relate them one first has to "raise one index" of a vertex turning it into a bilinear product.
This will also enable us to associate gauge algebras with non-Lagrangian equations.
For the self-dual Yang-Mills and gravity equations this construction previously appeared in  \cite{Monteiro:2011pc}.

 In the following we will use capital
Latin letters to denote collectively helicity, momentum and internal  labels
\begin{equation}
\label{6aug2}
\Phi^A \equiv\Phi^{\lambda|\alpha}(q), \qquad A\equiv \{\lambda,q,\alpha \}.
\end{equation}
We will also denote the space  spanned by off-shell fields $\Phi^A$ as  ${V}$.
In these terms the cubic Hamiltonian density $h_{A_1A_2A_3}$ is naturally a totally symmetric three-form
on $V$, while the cubic action can be written as\footnote{Contrary to our previous notation here it seems more convenient
 to include the momentum-conserving
delta-function in $h_{A_1A_2A_3}$, see (\ref{6aug5}).}
\begin{equation}
\label{6aug3}
S_3 = -\frac{g}{3!}h_{A_1A_2A_3}\Phi^{A_1}\Phi^{A_2}\Phi^{A_3}.
\end{equation}

There is a natural inner product on ${ V}$ defined by 
\begin{equation}
\label{6aug4}
 (\Phi_1,\Phi_2) \equiv 
\sum_{\{\lambda_1|\alpha_1\}} \sum_{\{ \lambda_2|\alpha_2\}} \int d^4 q_1 d^4 q_2 \delta(q_1+q_2)\delta^{\alpha_1,\alpha_2}\delta^{\lambda_1+\lambda_2,0} \Phi^{\lambda_1|\alpha_1}_1(q_1) \Phi^{\lambda_2|\alpha_2}_2(q_2).
\end{equation}
It is non-degenerate and can be used to raise/lower indices, thus identifying ${ V}$ and its dual  $V^*$ in the
 usual way. As it is not hard to see from 
(\ref{6aug4}), raising/lowering an index  results into  exchange of helicity and momentum labels
to opposite ones, while keeping internal indices intact.
One can rephrase this by saying that contraction with the inner form (\ref{6aug4}) swaps ingoing  for outgoing particles and vice versa.
Using the inner product (\ref{6aug4}) the free action can be written as
\begin{equation}
\label{6aug04}
S_2 = \frac{1}{2}(\Phi,\Box\Phi).
\end{equation}

For definiteness, in the following we will  consider chiral cubic theories with  antiholomorphic vertices.
Then, we have
\begin{equation}
\label{6aug5}
h^{A_1}{}_{A_2A_3} =- C^{-\lambda_1|\alpha_1,\lambda_2|\alpha_2,\lambda_3|\alpha_3}\delta(-q_1+q_2+q_3) \frac{\bar{\mathbb{P}}^{-\lambda_1+\lambda_2+\lambda_3}_{23}}{(-\beta_1)^{-\lambda_1}\beta_2^{\lambda_2}\beta_3^{\lambda_3}}.
\end{equation}
In these terms the field equations  read
\begin{equation}
\label{6aug6}
\Box_1\Phi^{A_1} -\frac{g}{2}h^{A_1}{}_{A_2A_3}\Phi^{A_2}\Phi^{A_3}=0.
\end{equation}

Finally, we define the structure constants as
\begin{equation}
\label{6aug7}
f^{A_1}{}_{A_2A_3}\equiv \frac{1}{2}h^{A_1}{}_{A_2A_3}\frac{\beta_2\beta_3}{\beta_1 \bar{\mathbb{P}}_{23}}.
\end{equation}
Clearly, $f^{A_1}{}_{A_2A_3}$ is antisymmetric in its two lower indices as a consequence of total symmetry of $h_{A_1A_2A_3}$.
As it will be demonstrated below, in all relevant cases $f^{A_1}{}_{A_2A_3}$ satisfies the Jacobi identity. This implies that
each chiral cubic theory defines a Lie algebra with a commutator 
\begin{equation}
\label{6aug8}
[T_{A_2},T_{A_3}] \equiv f^{A_{1}}{}_{A_2A_3}T_{A_1},
\end{equation}
where $T_{A_i}$ are the basis elements of $V$ playing the role of Lie algebra generators. We will call
the Lie algebra defined by structure constants (\ref{6aug7}) the {\emph{gauge algebra}} of the associated
chiral cubic theory.

The fact that each chiral cubic  theory is associated with some Lie algebra is, of
course, not surprising. 
As it will be clarified below, it is just another way to say that chiral  cubic 
theories are  some versions of the self-dual Yang-Mills theory with properly identified gauge algebras.
For parity invariant theories a similar idea underlies Cartan's approach to gravity and
the frame-like approach 
to higher spin theories \cite{Vasiliev:1980as,Fradkin:1987ks,Fradkin:1986qy}. Let us stress, however, 
that in the light-cone approach the gauge symmetry is completely fixed, so to relate the algebra defined above
to the gauge symmetry of the self-dual Yang-Mills equations one should first undo the light-cone gauge fixing.
This will be explained in details in section \ref{Sect5}.

It is often more convenient to deal not with the Lie algebra generators, but with the coefficients of Lie algebra
elements. Let 
\begin{equation}
\label{6aug08}
E_1 = [E_2,E_3]
\end{equation}
and $E^{A_i}_i$ be the coefficients of Lie algebra elements $E_i$ with respect to a basis $T_A$. Then, (\ref{6aug8})
implies 
\begin{equation}
\label{6aug008}
E_1^{A_1} = f^{A_1}{}_{A_2A_3}E_2^{A_2}E_3^{A_3}.
\end{equation}

In the following we will also discuss non-Lagrangian equations of motion. In this case one can extract the 
gauge algebra structure constants directly from field equations using (\ref{6aug6}), (\ref{6aug7}).
In the remainder of this section we will demonstrate for a list of theories that the structure constants that we
defined here, indeed, satisfy the Jacobi identity.

\subsection{Self-dual Yang-Mills theory}
\label{sdym}
It follows from the discussion of  section \ref{caht} that via truncating any consistent
parity invariant massless theory to a single set of cubic vertices one should obtain a
consistent chiral cubic  theory. Applying this conclusion to Yang-Mills theory we find
the self-dual Yang-Mills action as it was given by Chalmers and Siegel
 \cite{Chalmers:1996rq}
\begin{align}
\notag
S&= \int dq_1 dq_2 \delta(q_1+q_2) \Box_2\Phi^{-1|\alpha}(q_1)  \Phi^{1|\alpha}(q_2)\\
\label{8aug1}
& \qquad + \int dq_1dq_2 dq_3 \delta (q_1+q_2+q_3) g f^{\alpha_1\alpha_2\alpha_3} 
\frac{\bar{\mathbb{P}}_{23}}{\beta_2\beta_3}\beta_1 \Phi^{-1|\alpha_1}(q_1)
\Phi^{1|\alpha_2}(q_2)\Phi^{1|\alpha_3}(q_3),
\end{align}
where $f^{\alpha_1\alpha_2\alpha_3}$ are the structure constants of some internal Lie
algebra $\mathfrak{g}$.
By varying (\ref{8aug1}) with respect to $\Phi^{\alpha|-1}$ we obtain the self-dual Yang-Mills
equations in  light-cone gauge. In the coordinate representation this yields
\begin{equation}
\label{7aug01}
\Box \Phi^{\alpha_1|1} - 2g f^{\alpha_1\alpha_2\alpha_3} \partial^+ \left(\frac{\bar \partial}{\partial^+} \Phi^{\alpha_2|1} \Phi^{\alpha_3|1} \right)=0,
\end{equation}
where we used antisymmetry of $f^{\alpha_1\alpha_2\alpha_3}$ to simplify the result.

Equations (\ref{7aug01}) feature only the positive helicity field $\Phi^{\alpha|1}$ and, thus, form a closed 
non-Lagrangian sector of the Chalmers-Siegel theory.
By comparing (\ref{7aug01}) with (\ref{6aug6}) we can identify that
\begin{equation}
\label{7aug1}
h^{A_1}{}_{A_2A_3} = 2  f^{a_1a_2a_3} \delta(-q_1+q_2+q_3) \frac{\bar{\mathbb{P}}_{23}}{\beta_2\beta_3}\beta_1
\delta^{\lambda_1,1}\delta_{\lambda_2,1}\delta_{\lambda_3,1}
\end{equation}
and, accordingly, from (\ref{6aug7}) we obtain
\begin{equation}
\label{7aug02}
f^{A_1}{}_{A_2A_3} = f^{\alpha_1\alpha_2\alpha_3} \delta(-q_1+q_2+q_3) \delta^{\lambda_1,1}\delta_{\lambda_2,1}\delta_{\lambda_3,1}.
\end{equation}
It is not hard to see that the Jacobi identity for $f^{A_1}{}_{A_2A_3}$ is satisfied as a consequence of the Jacobi identity
for $f^{\alpha_1\alpha_2\alpha_3}$. Let us denote the algebra defined by (\ref{7aug02}) as
$\mathfrak{G}_{\rm YM}$.

To understand how this Lie algebra acts, we consider the commutator (\ref{6aug08}) in 
components. For the coefficients of the Lie algebra
elements featuring this commutator we have
\begin{equation}
\label{7aug002}
E_1^{1|\alpha_1}(q_1)=\int d^4q_2d^4 q_3\delta(-q_1+q_2+q_3) f^{\alpha_1\alpha_2\alpha_3}E_2^{1|\alpha_2}(q_2)
E_3^{1|\alpha_3}(q_3).
\end{equation}
By making the Fourier transform back to space-time variables one finds 
\begin{equation}
\label{7aug0002}
E_1^{1|\alpha_1}(x) = f^{\alpha_1\alpha_2\alpha_3}E_2^{1|\alpha_2}(x)E_3^{1|\alpha_3}(x).
\end{equation}
In other words, $\mathfrak{G}_{\rm YM}$ is a Lie algebra of $\mathfrak{g}$-valued functions in space-time
with a pointwise commutator
\begin{equation}
\label{7aug00002}
\mathfrak{G}_{\rm YM} = \mathfrak{g}\otimes C(x^-,x^+,x,\bar x).
\end{equation}

Let us now come back to the action (\ref{8aug1}) and vary it with respect to $\Phi^{\alpha|1}$.
The resulting equation of motion describes propagation of a negative helicity field $\Phi^{\alpha|-1}$ in a self-dual background
 given by (\ref{7aug01}). Rewriting the Hamiltonian density in the manifestly symmetric form
\begin{align}
\notag
h_{A_1A_2A_3} &= -2 f^{\alpha_1\alpha_2\alpha_3}\delta(q_1+q_2+q_3)
 \Big( \frac{\bar{\mathbb{P}}_{23}}{\beta_2\beta_3}\beta_1 \delta_{\lambda_1,-1}\delta_{\lambda_2,1}\delta_{\lambda_3,1}\\
 \label{7aug3}
 &\qquad\qquad\qquad\qquad  +
 \frac{\bar{\mathbb{P}}_{31}}{\beta_3\beta_1}\beta_2 \delta_{\lambda_2,-1}\delta_{\lambda_3,1}\delta_{\lambda_1,1}
 +\frac{\bar{\mathbb{P}}_{12}}{\beta_1\beta_2}\beta_3 \delta_{\lambda_3,-1}\delta_{\lambda_1,1}\delta_{\lambda_2,1}
  \Big),
\end{align}
we easily find the associated structure constants 
\begin{align}
\notag
f^{A_1}{}_{A_2A_3}& = f^{\alpha_1\alpha_2\alpha_3} \delta(-q_1+q_2+q_3)\big( \delta_{\lambda_1,1}\delta_{\lambda_2,1}\delta_{\lambda_3,1}
\\
 &\qquad\qquad\qquad \qquad +\left(\frac{\beta_2}{\beta_1} \right)^2\delta_{\lambda_1,-1}\delta_{\lambda_2,-1}\delta_{\lambda_3,1}
+\left(\frac{\beta_3}{\beta_1} \right)^2 \delta_{\lambda_1,-1}\delta_{\lambda_2,1}\delta_{\lambda_3,-1}\big).
\label{7aug4}
\end{align}
Again, it is not hard to check that the Jacobi identity is satisfied. Let us denote the associated algebra as
$\mathfrak{G}^{\rm L}_{\rm YM}$. Here and below label "${\rm L}$" refers to the fact that the algebra is 
associated with Lagrangian equations of motion.

To understand better how this algebra acts let us again look at the commutator (\ref{6aug08}) in components. Now to define a Lie algebra
element of $\mathfrak{G}^{\rm L}_{\rm YM}$ we need two $\mathfrak{g}$-valued functions in  space-time: $E^{1|\alpha}(q)$ and $E^{-1|\alpha}(q)$.
It is convenient to make a rescaling 
\begin{equation}
\label{7aug5}
\tilde E^{1|\alpha}(q)\equiv  E^{1|\alpha}(q), \qquad \tilde E^{-1|\alpha}(q)\equiv  \beta^2E^{-1|\alpha}(q).
\end{equation}
After performing the Fourier transform for rescaled algebra generator's coefficients we find 
\begin{align}
\notag
\tilde E_1^{1|\alpha_1}(x)& = f^{\alpha_1\alpha_2\alpha_3}\tilde E_2^{1|\alpha_2}(x)\tilde E_3^{1|\alpha_3}(x),\\
\label{7aug6}
\tilde E_1^{-1|\alpha_1}(x) & = f^{\alpha_1\alpha_2\alpha_3}\tilde E_2^{-1|\alpha_2}(x)\tilde E_3^{1|\alpha_3}(x)+
f^{\alpha_1\alpha_2\alpha_3}\tilde E_2^{1|\alpha_2}(x)\tilde E_3^{-1|\alpha_3}(x).
\end{align}
Formally, one can regard $\mathfrak{G}^{\rm L}_{\rm YM}$ as an algebra of functions of one Grassmann variable $\theta$
taking values in $\mathfrak{G}_{\rm YM}$, that is 
\begin{equation}
\label{24aug1}
\mathfrak{G}^{\rm L}_{\rm YM} = \mathfrak{G}_{\rm YM}\otimes C[\theta].
\end{equation}
Indeed, given $\mathfrak{G}_{\rm YM}$ generated by $T^1$, one can construct $\mathfrak{G}^{\rm L}_{\rm YM}$
by adding generators 
\begin{equation*}
T^{-1}\equiv T^1 \theta.
\end{equation*}
 Then, recalling that for Grassmann variables $\theta^2=0$, one easily recovers
 commutation relations (\ref{7aug6}).

\subsection{Self-dual gravity}
\label{sect32}
Similarly to the case of the Yang-Mills theory, the Einstein-Hilbert action can be consistently
truncated to a single cubic vertex, resulting in the self-dual gravity action \cite{Chalmers:1996rq}
\begin{align}
\notag
S&= \int dq_1 dq_2 \delta(q_1+q_2) \Box_2\Phi^{-2}(q_1)  \Phi^{2}(q_2)\\
\label{24aug2}
& \qquad\qquad \qquad \qquad  + \int dq_1dq_2 dq_3 \delta (q_1+q_2+q_3) g 
\frac{\bar{\mathbb{P}}^2_{23}}{\beta^2_2\beta^2_3}\beta^2_1 \Phi^{-2}(q_1)
\Phi^{2}(q_2)\Phi^{2}(q_3).
\end{align}
 By varying it with respect to $\Phi^{-2}$ we obtain 
the self-dual gravity equations
\begin{equation}
\label{24aug2e}
\Box \Phi^{2} + 2g  \left(\partial^+\right)^2 \left(\frac{\bar \partial^2}{\left(\partial^+\right)^2} \Phi^{2} \Phi^{2}
- \frac{\bar \partial}{\partial^+} \Phi^{2} \frac{\bar \partial}{\partial^+}  \Phi^{2} \right)=0.
\end{equation}

Comparing this equation with  (\ref{6aug6}), (\ref{6aug7}) we find that for  self-dual gravity 
\begin{equation}
\label{24aug3}
f^{A_1}{}_{A_2A_3} =  \delta(-q_1+q_2+q_3) \frac{\bar{\mathbb{P}}_{23}}{\beta_2\beta_3}\beta_1 \delta^{\lambda_1,2}\delta_{\lambda_2,2}\delta_{\lambda_3,2}.
\end{equation}
One way to see that these structure constants satisfy the Jacobi identity  is as follows.
The Jacobi identity reads
\begin{equation}
\label{27aug5}
f^{A_4}{}_{A_3 E} f^E{}_{A_1A_2}+f^{A_4}{}_{A_2 E} f^E{}_{A_3A_1}+f^{A_4}{}_{A_1 E} f^E{}_{A_2A_3}=0.
\end{equation}
Noticing that all $\beta$-dependence trivially cancels, we are left with
\begin{equation}
\label{27aug6}
\bar{\mathbb{P}}_{43}   \bar{\mathbb{P}}_{12} + \bar{\mathbb{P}}_{42} \bar{\mathbb{P}}_{31} +  \bar{\mathbb{P}}_{41} \bar{\mathbb{P}}_{23}=0,
\end{equation}
which is just another form of  the well-known Schouten identity, see (\ref{27aug7}).

For later it will be useful to introduce the following basis for $\bar{\mathbb{P}}_{ij}$ with $i,j =1,2,3,4$
\begin{align}
\notag
2A & \equiv \bar{\mathbb{P}}_{12}+\bar{\mathbb{P}}_{34}=-\bar{\mathbb{P}}_{14}+\bar{\mathbb{P}}_{23},\\
\notag
2B & \equiv \bar{\mathbb{P}}_{13}-\bar{\mathbb{P}}_{24}=\bar{\mathbb{P}}_{34}-\bar{\mathbb{P}}_{12},\\
\label{27aug8}
2C & \equiv \bar{\mathbb{P}}_{14}+\bar{\mathbb{P}}_{23}=-\bar{\mathbb{P}}_{13}-\bar{\mathbb{P}}_{24}.
\end{align}
In these terms (\ref{27aug6}) becomes a simple identity
\begin{equation}
\label{27aug9}
B^2 -A^2 + C^2 - B^2 + A^2 - C^2=0.
\end{equation}

Let us clarify what is the algebra  that these structure constants define. As before, it is more instructive
to look at coefficient of the Lie algebra elements entering the commutator. To start, we 
rescale them as
\begin{equation}
\label{24aug4}
\tilde E^2(q) \equiv \beta^{-1} E^2(q).
\end{equation}
Then, after performing the inverse Fourier transform 
for the commutator we obtain
\begin{equation}
\label{24aug5}
\tilde E_1^{2}(x) = 
(\bar\partial_2 \partial^+_3- \partial^+_2  \bar\partial_3) E_2^{2}(x) E_3^{2}(x) \equiv
\bar\partial \tilde E_2^{2}(x)\partial^+\tilde E_3^{2}(x)- \partial^+  \tilde E_2^{2}(x)\bar\partial \tilde E_3^{2}(x).
\end{equation}
The operator acting on the right hand side is nothing but the Poisson bracket in the 2d plane with coordinates $(x^-,x)$
and the standard symplectic form.
Thus, the Lie algebra defined by (\ref{24aug5}) is an algebra of functions of $(x^+,\bar x)$ 
valued in the algebra of 2d area-preserving diffeomorphisms with a point-wise commutator, which will be denoted as
\begin{equation}
\label{24aug6}
\mathfrak{G}_{\rm GR} = {\rm SDiff}(x^-,x)\otimes C(x^+,\bar x).
\end{equation}
In the present form the connection of self-dual gravity to area-preserving diffeomorphisms was identified in \cite{Monteiro:2011pc}.

Next we come back to the action (\ref{24aug2}) and consider all equations together. Acting as in the previous 
section we find that they determine the following  structure constants 
\begin{align}
\notag
f^{A_1}{}_{A_2A_3}& = f^ \delta(-q_1+q_2+q_3)\frac{\bar{\mathbb{P}}_{23}}{\beta_2\beta_3}\beta_1\big( \delta_{\lambda_1,2}\delta_{\lambda_2,2}\delta_{\lambda_3,2}
\\
 &\qquad\qquad\qquad \qquad +\left(\frac{\beta_2}{\beta_1} \right)^4\delta_{\lambda_1,-2}\delta_{\lambda_2,-2}\delta_{\lambda_3,2}
+\left(\frac{\beta_3}{\beta_1} \right)^4 \delta_{\lambda_1,-2}\delta_{\lambda_2,2}\delta_{\lambda_3,-2}\big),
\label{25aug1}
\end{align}
which, in turn, define the Lie algebra that can be obtained by extending $\mathfrak{G}_{\rm GR}$ with an extra
Grassmann variable as in (\ref{24aug1})
\begin{equation}
\label{25aug2}
\mathfrak{G}^{\rm L}_{\rm GR} = \mathfrak{G}_{\rm GR}\otimes C[\theta].
\end{equation}

\subsection{Chiral higher spin theory}
\label{section3.3}
In \cite{Metsaev:1991mt,Metsaev:1991nb} consistency conditions (\ref{2aug8}), (\ref{2aug10}) were solved in the 
higher spin case. For colour-neutral higher spin fields the solution of (\ref{2aug8}) is
\begin{equation}
\label{27aug1}
C^{\lambda_1,\lambda_2,\lambda_3}= 2 \frac{\ell^{\lambda_1+\lambda_2+\lambda_3-1}}{(\lambda_1+\lambda_2+\lambda_3-1)!},
\end{equation}
where $\ell$ is an additional dimensionful coupling constant necessary to compensate for dimensions carried
by derivatives. A systematic derivation of (\ref{27aug1}) was given in  \cite{Ponomarev:2016lrm}.

As it was already reviewed in section \ref{sect24}, one can truncate the light-cone deformation procedure by 
setting $\bar C$ and all higher vertices to zero. This results in the chiral higher spin theory 
with the action \cite{Ponomarev:2016lrm}\footnote{A chiral version of the Vasiliev equations \cite{Vasiliev:1990en,Vasiliev:2003ev} was proposed in \cite{Iazeolla:2007wt}. 
It remains to be understood whether this construction is related to the chiral higher spin theory discussed in the present paper.}
\begin{align}
\notag
&S =\frac{1}{2} \sum_{\lambda_1,\lambda_2} \int d^4 q_1 d^4 q_2 \delta(q_1+q_2)\delta^{\lambda_1+\lambda_2,0} 
\Box_2\Phi^{\lambda_1}_1\Phi^{\lambda_2}_2\\
\label{27aug2}
&\;+\frac{g}{3}\sum_{\lambda_1,\lambda_2,\lambda_3}\int d^4 q_1 d^4 q_2 d^4 q_3 \delta(q_1+q_2+q_3)
\frac{\ell^{\lambda_1+\lambda_2+\lambda_3-1}}{(\lambda_1+\lambda_2+\lambda_3-1)!} \frac{\bar{\mathbb{P}}^{\lambda_1+\lambda_2+\lambda_3}_{23}}{\beta_1^{\lambda_1}\beta_2^{\lambda_2}\beta_3^{\lambda_3}}
\Phi_1^{\lambda_1}\Phi_2^{\lambda_2}\Phi_3^{\lambda_3}.
\end{align}
Contrary to the lower spin cases we cannot present it as a cubic truncation of some parity invariant theory
as the latter theory is not yet known. However, striking similarities between the pattern followed by lower spin
theories and the pattern that we will exhibit below  suggest that a consistent parity invariant
completion of (\ref{27aug2}) might exist.

Structure constants associated via (\ref{6aug5}) with this chiral cubic  theory are
\begin{equation}
\label{27aug3}
f^{A_1}{}_{A_2A_3} = \delta (-q_1+q_2+q_3) 
\frac{(\ell\; \bar{\mathbb{P}}_{23})^{-\check\lambda_1+\check\lambda_2+\check\lambda_3}}{(-\check\lambda_1+\check\lambda_2+\check\lambda_3)!
(-\beta_1)^{-\check\lambda_1}\beta_2^{\check\lambda_2}\beta_3^{\check\lambda_3}},
\end{equation}
where for brevity we introduced
\begin{equation}
\label{27aug4}
\check \lambda_i \equiv \lambda_i -1.
\end{equation}
To verify the Jacobi identity we first note that $\beta$ dependence trivially cancels out.
What remains to check is 
\begin{align}
\label{27aug10}
\sum_{\lambda_E} \frac{(\ell \; \bar{\mathbb{P}}_{34})^{-\check\lambda_4+\check\lambda_3+\check\lambda_E}}{(-\check\lambda_4+\check\lambda_3+\check\lambda_E)!}
\frac{(\ell \; \bar{\mathbb{P}}_{12})^{-\check\lambda_E+\check\lambda_1+\check\lambda_2}}{(-\check\lambda_E+\check\lambda_1+\check\lambda_2)!}+
\text{2 terms}=0,
\end{align}
where "$E$" refers to the exchanged field, while the implicit two terms are obtained by cyclic permutations in $(1,2,3)$ from the first one, see (\ref{27aug5}).
Recalling that only terms with the total helicity being even contribute, we find that (\ref{27aug10}) reduces to a trivial identity
\begin{equation}
\label{27aug11}
(\ell\;  B)^\Lambda- (\ell \; A)^\Lambda+(\ell\; C)^\Lambda-(\ell\;  B)^\Lambda+(\ell\;  A)^\Lambda-(\ell\;  C)^\Lambda=0,
\end{equation}
where
\begin{equation}
\label{27aug12}
\Lambda \equiv -\check \lambda_4 + \check \lambda_1+\check \lambda_2+\check \lambda_3
\end{equation}
and $A$, $B$ and $C$ were defined in (\ref{27aug8}).

Let us now rewrite the commutator defined by (\ref{27aug3}) in a more convenient form. As before, we scale away $\beta$ dependence
by defining 
\begin{equation}
\label{27aug13}
\tilde E^\lambda(q) \equiv \beta^{-\check \lambda} E^\lambda(q).
\end{equation}
Next we combine coefficient for different helicities into a single generating function as
\begin{equation}
\label{27aug14}
\tilde E(q;z)\equiv \sum_{\lambda=-\infty}^\infty \tilde E^\lambda(q) z^{\check \lambda}.
\end{equation}
Let us stress that the sum goes from minus to plus infinity, so (\ref{27aug14}) is the Laurent series in $z$ at $z=0$.
Finally, by making the inverse Fourier transform, we obtain that for $E_1=[E_2,E_3]$ the generating functions for the respective
components satisfy
\begin{equation}
\label{27aug15}
\tilde E_1(x;z) =\sinh \left( \frac{\ell}{z} \left(\bar\partial_2 \partial^+_3- \partial^+_2  \bar\partial_3\right) \right)    \tilde E_2(x;z) \tilde E_3(x;z).
\end{equation}

Similarly to the case of gravity the commutator involves only derivatives with respect to $x^-$ and $x$, while
$x^+$ and $\bar x$   dependence enters trivially. Hence, the algebra defined by (\ref{27aug15}) has the structure 
\begin{equation}
\label{27aug16}
\mathfrak{G}^{\rm L}_{\rm HS} = {\rm hs}(x^-,x)\otimes C(x^+,\bar x),
\end{equation}
where ${\rm hs}$ is defined by the same formula (\ref{27aug15})  as $\mathfrak{G}^{\rm L}_{\rm HS}$, but for $x^+$- and $\bar x$-independent coefficient functions and can be
regarded as a higher spin generalisation of area-preserving diffeomorphisms ${\rm SDiff}(x^-,x)$. 

Note, 
that contrary to lower spin cases the higher spin structure constants (\ref{27aug3}) were immediately derived 
from the action, that is they are immediately associated with a symmetric $h_{A_1A_2A_3}$ and do not
require any Lagrangian completion as in (\ref{24aug1}), (\ref{25aug2}). This is why we immediately 
supplemented $\mathfrak{G}_{\rm HS}$ with "${\rm L}$" label. The question whether this algebra 
admits any subalgebras similar to $\mathfrak{G}_{\rm YM}$ and $\mathfrak{G}_{\rm GR}$ will be discussed in section \ref{subalgebras}.

Another interesting observation comes from comparison of qualitative features of the algebra defined by (\ref{27aug15})
and the higher spin algebra in AdS, the Eastwood-Vasiliev algebra \cite{Eastwood:2002su,Vasiliev:2003ev}\footnote{For earlier results on higher spin
algebras in 4d AdS space, see \cite{Fradkin:1986ka,Vasiliev:1986qx}.}. It is straightforward to see that the commutator of two generators of fixed helicities
computed with (\ref{27aug15}) is schematically of the form 
\begin{equation}
\label{27aug17}
[T_{\lambda_1},T_{\lambda_2}] = T_{\lambda_1+\lambda_2-2}+ T_{\lambda_1+\lambda_2-4}+\dots, 
\end{equation}
where the sum goes all the way down to minus infinity. In particular, 
\begin{equation}
\label{27aug18}
[T_{\lambda},T_{2}] = T_{\lambda}+ T_{\lambda-2}+\dots
\end{equation}
with an infinite tail of terms with arbitrarily large negative helicities. In contrast, for the 
Eastwood-Vasiliev algebra one has
\begin{equation}
\label{27aug19}
[T_{s_1},T_{s_2}] = T_{s_1+s_2-2}+ T_{s_1+s_2-4}+\dots + T_{|s_1-s_2|+2} 
\end{equation}
and, hence, 
\begin{equation}
\label{27aug20}
[T_s,T_{2}] = T_s.
\end{equation}
The reason why the sum in (\ref{27aug19}) has spin limited from below is that the Eastwood-Vasiliev
bracket is constructed in terms of tensor contractions and the sum truncates whenever indices of
tensors are saturated. The bracket (\ref{27aug15}), though, looks similar, does not imply any
tensor contractions, which eventually results in infinite sums (\ref{27aug17}), (\ref{27aug18}).
Similar phenomenon occurs in three dimensions, where higher spin generator also do not
carry any tensor indices and, thus, one can formally consider negative spin generators and 
algebras involving them \cite{Pope:1989sr}.

\subsection{Coloured chiral higher spin theory}
\label{sect34}
One can also consider a setup where higher spin fields additionally carry some internal labels. The 
associated consistency condition (\ref{2aug8}) was studied  
and two solutions were found \cite{Metsaev:1991nb,Metsaev:1991thesis}. In a more restricted setup where higher spin fields take values in some
Lie algebra and cubic vertices are explicitly proportional to the Lie algebra structure constants a
systematic way to solve (\ref{2aug8}) was given in \cite{Ponomarev:2016lrm}. It results in 
\begin{equation}
\label{28aug1}
C^{\lambda_1|\alpha_1,\lambda_2|\alpha_2,\lambda_3|\alpha_3}= 2 f^{\alpha_1\alpha_2\alpha_3} \delta(\lambda_1+\lambda_2+\lambda_3-1),
\end{equation}
which leads to the coloured chiral higher spin theory given by the action 
\begin{align}
\notag
&S =\frac{1}{2} \sum_{\lambda_1,\lambda_2} \int d^4 q_1 d^4 q_2 \delta(q_1+q_2)\delta^{\lambda_1+\lambda_2,0} 
\Box_2\Phi^{\lambda_1|\alpha}_1\Phi^{\lambda_2|\alpha}_2\\
\notag
&\qquad \qquad+\frac{g}{3}\sum_{\lambda_1,\lambda_2,\lambda_3}\int d^4 q_1 d^4 q_2 d^4 q_3 \delta(q_1+q_2+q_3)
 \delta(\lambda_1+\lambda_2+\lambda_3-1)\\
 \label{28aug2}
& \qquad \qquad   \qquad \qquad \qquad \qquad \qquad \qquad \cdot f^{\alpha_1\alpha_2\alpha_3} \frac{\bar{\mathbb{P}}_{23}}{\beta_1^{\lambda_1}\beta_2^{\lambda_2}\beta_3^{\lambda_3}}
\Phi_1^{\lambda_1|\alpha_1}\Phi_2^{\lambda_2|\alpha_2}\Phi_3^{\lambda_3|\alpha_3}.
\end{align}

The associated structure constants (\ref{6aug7}) are given by 
\begin{equation}
\label{28aug3}
f^{A_1}{}_{A_2A_3} = f^{\alpha_1\alpha_2\alpha_3} \delta(-q_1+q_2+q_3)\frac{ \delta (-\check\lambda_1+\check\lambda_2+\check\lambda_3)}{(-\beta_1)^{-\check\lambda_1}
\beta_2^{\check \lambda_2}\beta_3^{\check\lambda_3}}.
\end{equation}
After rescaling the coefficient functions as in  (\ref{27aug13}) and combining them into a generating function (\ref{27aug14}),
 they define the commutator
\begin{equation}
\label{28aug4}
\tilde E^{\alpha_1}_1(x;z) =  f^{\alpha_1\alpha_2\alpha_3} \tilde E^{\alpha_2}_2(x;z) \tilde E^{\alpha_3}_3(x;z).
\end{equation}
This algebra, which will be denoted $\mathfrak{G}^{\rm L}_{\rm CHS}$, obviously, is just a loop algebra of that found for the self-dual Yang-Mills equations
in section  \ref{sdym}
\begin{equation}
\label{28aug5}
\mathfrak{G}^{\rm L}_{\rm CHS}= \mathfrak{G}_{\rm YM}\otimes C[z,z^{-1}].
\end{equation}
Similarly to (\ref{27aug16}) we will separate in this algebra  the $x^+$- and $\bar x$- independent 
part and denote it 
\begin{equation}
\label{28aug6}
{\rm chs}(x^-,x)= \mathfrak{g}\otimes C(x^-,x)\otimes C[z,z^{-1}].
\end{equation}

\subsection{Poisson chiral higher spin theory}
\label{sect3.5}
The algebra found in section \ref{section3.3} with a Lie bracket, see (\ref{27aug15})
\begin{equation}
\label{29aug1}
[  \tilde E_1(x;z), \tilde E_2(x;z)]\equiv  \sinh \left( \frac{\ell}{z} \left(\bar\partial_1 \partial^+_2- \partial^+_1  \bar\partial_2\right) \right)    \tilde E_1(x;z) \tilde E_2(x;z)
\end{equation}
is, clearly, very similar to the bracket constructed via the star-product. The latter bracket 
has a very well-known Poisson contraction. It is straightforward to make an analogous 
contraction of bracket (\ref{29aug1}), which leads to
\begin{equation}
\label{29aug2}
[  \tilde E_1(x;z), \tilde E_2(x;z)]_{\rm P}\equiv   \frac{\ell}{z} \left(\bar\partial_1 \partial^+_2- \partial^+_1  \bar\partial_2\right)    \tilde E_1(x;z) \tilde E_2(x;z).
\end{equation}
One can easily check that this bracket satisfies the Jacobi identity. Moreover, the power of $z$ in front of the bracket is fixed by 
the requirement of Lorentz invariance, see (\ref{1aug16}). 
We will denote the associated gauge algebra $\mathfrak{G}^{\rm L}_{\rm PHS}$ and its 
$x^+$- and $\bar x$-independent part by phs$(x^-,x)$.

Once the algebra is known one can go through our typical manipulation in the opposite direction
to find that it corresponds to the coupling constants
\begin{equation}
\label{29aug3}
C^{\lambda_1,\lambda_2,\lambda_3}= 2 \ell \delta (\lambda_1+\lambda_2+\lambda_3-2)
\end{equation}
and, hence, to the action
\begin{align}
\notag
&S =\frac{1}{2} \sum_{\lambda_1,\lambda_2} \int d^4 q_1 d^4 q_2 \delta(q_1+q_2)\delta^{\lambda_1+\lambda_2,0} 
\Box_2\Phi^{\lambda_1}_1\Phi^{\lambda_2}_2\\
\label{29aug4}
&\;+\frac{g}{3}\sum_{\lambda_1,\lambda_2,\lambda_3}\int d^4 q_1 d^4 q_2 d^4 q_3 \delta(q_1+q_2+q_3)
\ell \delta (\lambda_1+\lambda_2+\lambda_3-2)\frac{\bar{\mathbb{P}}^{2}_{23}}{\beta_1^{\lambda_1}\beta_2^{\lambda_2}\beta_3^{\lambda_3}}
\Phi_1^{\lambda_1}\Phi_2^{\lambda_2}\Phi_3^{\lambda_3}.
\end{align}
It is straightforward to see that consistency conditions (\ref{2aug8}) for this theory are satisfied as a consequence of those
for the chiral higher spin theory that was contracted. We will call the theory given by (\ref{29aug4}) the Poisson chiral higher spin theory.

Let us briefly discuss some properties of this theory and of the associated algebra (\ref{29aug2}). First of all, the 
action (\ref{29aug4}) contains only two-derivative terms. As a consequence, if we consider the associated parity invariant
theory, the quartic vertex needed to compensate $[H_3,J_3]$ and $[H_3,\bar J_3]$ contributions to (\ref{2aug7}) and (\ref{2aug9})
should also contain only two derivatives\footnote{To be more precise, it should have homogeneity degree one on each momentum $q$ and $\bar q$,
but, clearly, cannot be polynomial because of local obstructions.}. The same pattern continues to all orders. Of course, it is not clear at this point
how to resolve obstructions related to locality. Nevertheless, we can see already now that if the parity invariant completion
of (\ref{29aug4}) exists it will be very reminiscent of usual gravity.

Yet another attractive feature of the Poisson-contracted chiral higher spin theory is how its gauge algebra acts. The commutation relations
exhibit the following schematic  pattern 
\begin{equation}
\label{29aug5}
[T_{\lambda_1},T_{\lambda_2}] = T_{\lambda_1+\lambda_2-2}, 
\end{equation}
which leads to 
\begin{equation}
\label{29aug6}
[T_{\lambda},T_2] = T_{\lambda}.
\end{equation}
The last formula implies that the helicity-$\lambda$ generators form a representation of the Lie algebra generated
by $T_2$. A related property is necessary to identify higher spin fields with representations of the Poincare algebra.
In contrast, this feature is spoiled in (\ref{27aug18}) by subleading 
corrections.

It is also interesting to contrast the above discussion  with the situation in AdS.
There the Poisson contraction of the Eastwood-Vasiliev algebra is still well-defined. However, the typical
approach that one uses to construct the cubic action, the Fradkin-Vasiliev approach
\cite{Fradkin:1987ks,Fradkin:1986qy,Boulanger:2012dx}, requires that the gauge
algebra possesses a non-degenerate invariant form, which is no longer true for the contracted algebra.
This issue was extensively discussed recently in \cite{Boulanger:2013zza} and will be briefly reviewed in section \ref{sect3.7}.

\subsection{Non-Lagrangian subalgebras}
\label{subalgebras}
In sections \ref{section3.3}-\ref{sect3.5} we constructed three higher spin gauge algebras associated with
Lagrangian equations of motion via the procedure explained at the beginning of section \ref{sect3}.
 In this section we will show that they have some subalgebras, which can
be viewed as higher spin generalisations of algebras $\mathfrak{G}_{\rm YM}$ and $\mathfrak{G}_{\rm GR}$
found for the self-dual Yang-Mills and the self-dual gravity equations of motion.

Let us start from the simplest case of the coloured chiral higher spin theory with a gauge
algebra $\mathfrak{G}^{\rm L}_{\rm CHS}$. It is not hard to see that the commutator (\ref{28aug4})
preserves $\check \lambda$, see (\ref{27aug4}), which we will also call "level" and denote $l\equiv \check\lambda$.
In other words, for $E_3=[E_1,E_2]$ we have
\begin{equation}
\label{29aug7}
l_3=l_1+l_2.
\end{equation}

Clearly, if $l_1=l_2=0$ then we also have $l_3=0$. The associated subalgebra of $\mathfrak{G}^{\rm L}_{\rm CHS}$ is just $\mathfrak{G}_{\rm YM}$.
 For $l_1\ge 0$ and $l_2\ge 0$ we have $l_3\ge 0$, which can be naturally viewed as a higher spin
 counterpart of $\mathfrak{G}_{\rm YM}$.
Similarly,  $l \le 0$  defines a subalgebra of $\mathfrak{G}^{\rm L}_{\rm CHS}$. More generally,
it is not hard to see, that the condition on the spectrum $l\ge m$ defines a subalgebra of $\mathfrak{G}^{\rm L}_{\rm CHS}$
for any $m\ge 0$, while $l\le m$ defines a subalgebra of $\mathfrak{G}^{\rm L}_{\rm CHS}$ if $m \le 0$. 
This construction parallels a similar construction for diagonal, upper- and lower-triangular
subalgebras of matrix algebras.

In the same way, introducing the level as $l\equiv\lambda-2$ we find that it is conserved for the Poisson chiral higher spin theory.
This observation allows to construct Poisson analogues of all subalgebras just considered for the coloured case.
In particular, $l=0$ defines $\mathfrak{G}_{\rm GR}$ and $l \ge 0$ may be viewed as its higher spin generalisation.

Finally, introducing the level as $l\equiv\lambda-2$, we see that for the chiral higher spin theory we have
\begin{equation}
\label{29aug8}
l_3\le l_1+l_2.
\end{equation}
Note, that here we took into account that if internal degrees of freedom are absent, then the vertex
with $\lambda_1+\lambda_2+\lambda_3=1$ vanishes.
Equation (\ref{29aug8}) implies that out of constraints on the spectrum considered above only $l\le m$ with $m\le 0$
defines subalgebras of $\mathfrak{G}_{\rm HS}$.

Before closing this section we would like to stress that our intent here was not to classify all possible
subalgebras that $\mathfrak{G}^{\rm L}_{\rm HS}$, $\mathfrak{G}^{\rm L}_{\rm CHS}$ and $\mathfrak{G}^{\rm L}_{\rm PHS}$ possess.
Instead, we mainly wanted to highlight the most obvious non-Lagrangian subalgebras
 that can be viewed as higher spin generalisations of
$\mathfrak{G}_{\rm YM}$ and $\mathfrak{G}_{\rm GR}$ for the self-dual Yang-Mills and gravity equations. Some other subalgebras and contractions are briefly discussed in the following section.

\subsection{Other subalgebras and contractions}
\label{sect3.7}
First, let us note that, similarly to higher spin theories in AdS, the chiral higher spin theory can be truncated only to even helicities without breaking consistency. Similarly, one can see that the spectrum of the Poisson/coloured higher spin theory can be truncated to only even/odd helicities, leaving the theory consistent. As a consequence, similar relations hold for the associated higher spin gauge algebras. The spectra of these theories can be truncated even further.

For example, as was noted in \cite{Ponomarev:2016lrm}, a single higher spin field coupled minimally to gravity
defines a consistent theory. In other words, we put $C^{2,2,-2}=C^{2,\lambda,-\lambda}=\ell$, while other couplings are
vanishing. This corresponds to a non-Lagrangian gauge algebra with the only non-trivial commutators of the form
\begin{equation}
\label{29aug9}
[T_2,T_2]=T_2, \qquad [T_\lambda,T_2]=T_\lambda.
\end{equation}
A complete set of commutators for Lagrangian equations additionally contains
\begin{equation}
\label{29aug10}
[T_{-2},T_2]=T_{-2}, \qquad [T_{-\lambda},T_2]=T_{-\lambda}, \qquad [T_\lambda, T_{-\lambda}]= T_{-2}.
\end{equation}
Analogous arguments apply for a single higher spin field minimally coupled to  Yang-Mills theory.

Systematic classification of all consistent truncations of chiral theories   would require a detailed study of 
the consistency conditions (\ref{2aug8}) with a particular attention payed to various possibilities of setting coupling constants to zero in a coherent manner. Alternatively, one can  analyse the Jacobi identity for the associated gauge algebras and study consistent ways to set some of the structure constants to zero. A related analysis was carried out in \cite{Boulanger:2013zza}, where the Jacobi identity for higher spin algebras in AdS was solved in complete generality and it was found that the only two solutions are given by the Eastwood-Vasiliev algebra and its even spin subalgebra.

 It is worth to stress that an important role in that analysis was played by the so-called Fradkin-Vasiliev condition, which is the requirement that the resulting algebra has to posses a non-degenerate invariant inner product. This is necessary, in particular, to ensure that the free theory action, which is built by employing this inner product, is non-singular. This inner product should also be diagonal in spin, because this is what the structure of the free action implies.

 The Fradkin-Vasiliev condition leads to strong constraints on the higher spin algebra structure constants. In particular, it implies that
\begin{equation}
\label{9dec1}
f^s{}_{s2}\ne 0 \qquad \Rightarrow \qquad f^2{}_{ss}\ne 0,
\end{equation}
which can be seen by raising/lowering indices with the invariant metric. The left hand side of (\ref{9dec1}) is necessary to ensure that higher spin generators belong to appropriate representations of the isometry algebra, while the right hand side is satisfied for the Eastwood-Vasiliev algebra and not satisfied for its Poisson contraction. This example illustrates that the Fradkin-Vasiliev condition is a non-trivial additional requirement, that rules out otherwise consistent higher spin algebras.

 In fact, in proving uniqueness of the Eastwood-Vasiliev algebra, the Fradkin-Vasiliev condition was used quite extensively, see Appendix A of \cite{Boulanger:2013zza}. Some other examples of consistent Lie algebras violating the Fradkin-Vasiliev condition were also given there. In particular, it was noted that one can construct an algebra featuring only spin 2 and spin 3 generators with  non-zero commutators acting as in (\ref{29aug9}). More generally, we expect strong similarities between the structures of subalgebras/contractions of the chiral higher spin algebra (\ref{27aug15}) and the Eastwood-Vasiliev algebra with the Fradkin-Vasiliev condition relaxed. Indeed, the only essential qualitative difference between  these two algebras is that the generators of the chiral higher spin algebra are labelled by helicities, not spins, and, hence, these labels may take negative values. 
 
  A systematic study of subalgebras and consistent contractions of the chiral higher spin gauge algebra would require an  analysis similar to that of  Appendix A in \cite{Boulanger:2013zza}, but with the Fradkin-Vasiliev condition relaxed.  It will be considered elsewhere. Finally, it is worth to note, that in the light-cone approach, it is the separate treatment of negative and positive helicity fields  and the fact that the kinetic term (\ref{1aug1}) is non-diagonal in helicities, that allows to avoid the Fradkin-Vasiliev condition in its usual form and makes contractions, such as the Poisson contraction, consistent not only at the algebra level, but also at the level of action.

\subsection{Degenerate examples}
\label{sect3.8}
An even more trivial example of cubic chiral theory is provided by a spin one theory with a non-minimal $F^3$ interaction, where $F$
is the linearised field strength.
For this theory the only non-vanishing coupling
constant is  $C^{1,1,1}$ and the only non-vanishing commutator that this structure constant 
defines, schematically, is 
\begin{equation}
\label{29aug11}
[T_1,T_1]=T_{-1}.
\end{equation}
As $T_{-1}$ commutes with other generators, any nested commutator vanishes and the Jacobi identity
is trivially satisfied. Similar arguments apply to spin $2$ theory with $R^3$ interaction, where $R$
is the linearised Riemann tensor. 

So far we considered many examples of chiral cubic theories and found that following
the same simple prescription we were able to define structure constants that satisfy the Jacobi identity and
thus define a Lie algebra. There is another example of a cubic theory that we did not consider yet ---
a scalar theory with $\Phi^3$ interaction. This is not a gauge theory and we do not expect 
that it can be rewritten as the self-dual Yang-Mills theory.
 Thus, we can anticipate the prescription for a gauge algebra that we applied so far  to fail in this case. 
And, indeed, if we carry out the standard steps, we will, 
first of all, find that structure constants (\ref{6aug7}) are non-local. Moreover, the Jacobi
identity is not satisfied because
\begin{equation}
\label{29aug12}
\frac{1}{B^2-A^2}+ \frac{1}{C^2-B^2}+\frac{1}{A^2-C^2} \ne 0,
\end{equation}
see (\ref{27aug8}) for notations.

Having considered so many examples we were able to convince ourselves that each chiral cubic
theory is indeed related to some gauge algebra. It was, however, not clear what is the role played by these
algebras. In order to clarify this we will now consider two ways how these gauge algebras manifest themselves.
First, we will show that field equations for chiral cubic theories can be literally reformulated as the self-dual
Yang-Mills equations with gauge algebras replacing usual colour algebras. Then, we will show how the gauge algebras are connected to the colour-kinematics duality.

\section{Self-duality and hidden symmetries}
\label{Sect5}
In this section we will show that the structure of the Lagrangian for the cubic chiral theories that we identified
in section \ref{sect3} implies that these theories can be understood as the generalised self-dual Yang-Mills theories
with the usual colour algebra replaced by associated gauge algebras.
 To see that we will do a series of  manipulations, which are rather standard  in the context of the analysis  of the self-dual Yang-Mills
equations \cite{Mason:1991rf,Hitchin:1999at}. For the analysis that includes self-dual gravity in a uniform way, see also  \cite{Park:1989vq,Park:1990fp}.

Equations of motion for cubic chiral theories can be written as
\begin{equation}
\label{31aug12}
\Box \Phi - 2g  \partial^+ \left[\frac{\bar \partial}{\partial^+} \Phi ,\Phi \right]=0,
\end{equation}
where $[\cdot, \cdot]$ is the Lie bracket of the gauge algebra of the theory.
For higher spin theories we assume that all fields are combined into a single generating function $\Phi(x;z)$, see
(\ref{27aug14}),  after rescaling (\ref{27aug13}) was performed.
Next,  we denote 
\begin{equation}
\label{31aug13}
A^{x}\equiv \Phi
\end{equation}
and introduce another field $A^{-}$ via 
\begin{equation}
\label{31aug14}
\partial^+ A^{-}+ \partial^{\bar x} A^{x}=0.
\end{equation}
In these terms (\ref{31aug12}) can be rewritten as
\begin{equation}
\label{31aug15}
\partial^- A^{x} - \partial^x A^- + g[A^-,A^x]=0.
\end{equation}
We can recognise here the condition that sets to zero the $(+\bar x)$ component of the Yang-Mills curvature 
\begin{equation}
\label{31aug16}
F_{ij} = \partial_i A_j - \partial_j A_i +g [A_i,A_j].
\end{equation}

Finally, we notice that (\ref{31aug14}), (\ref{31aug15}) can be viewed as a result of the
light-cone gauge fixing $A^+=0$
 in a system of equations
\begin{align}
\notag
F^{-x}=&\;0,\\
\notag
F^{+-}+ F^{\bar x x}=&\;0,\\
F^{+\bar x}=&\;0,
\label{31aug18}
\end{align}
where  gauge symmetry acts in the standard way
\begin{equation}
\label{31aug019}
\delta A = d\epsilon +g [A,\epsilon].
\end{equation}
Indeed, the last equation in (\ref{31aug18}) implies that for each 2d plane with fixed $x^+$ and $\bar x$
the connection is pure gauge. Hence, we can use gauge freedom to set the components of the connection 
along these planes to zero, $A^+=A^{\bar x}=0$. Then the second equation in (\ref{31aug18}) reproduces
(\ref{31aug14}), while the first one is the same as (\ref{31aug15}). It remains to note that
(\ref{31aug18}) are the self-dual Yang-Mills equations with a properly defined Hodge star
\begin{equation}
\label{31aug19}
F = i* F.
\end{equation}

In other words, we have just shown that all cubic antiholomoprhic theories result into the generalised self-dual
Yang-Mills equations with a gauge algebra defined by structure constants (\ref{6aug7}). 
What prevents us from simply calling them  the self-dual Yang-Mills equations is that 
 gauge Lie algebras that appear for self-dual gravity and other chiral cubic  theories involve 
space-time derivatives. This means that, strictly speaking,  they cannot be regarded as internal symmetries.
This is also the reason why we cannot immediately attribute all properties of the self-dual Yang-Mills equations
to other theories without doing further analysis.

Before proceeding with hidden symmetries, let us make a couple of comments about  equations (\ref{31aug18}).
To start, we would like to emphasise that the procedure of undoing light-cone gauge that lead us to (\ref{31aug18})
is, clearly,  not  opposite to the light-cone gauge fixing that one has to perform with the Fronsdal theory  \cite{Fronsdal:1978rb}
to obtain the action we 
started from in section \ref{caht}. Indeed, in (\ref{31aug18}) every helicity is represented by a  four-component field 
$A_i$, not by a rank-$s$ symmetric double-traceless tensor as in the Fronsdal approach. 

So far index ``$i$'' of $A_i$ was just labeling its components and we were not specific whether this notation implies any transformation properties with respect to the Lorentz group. Let us now clarify this point.
First, we note, that the only requirement that we actually need to demand is that $A^x \equiv \Phi$ transforms as was defined originally in (\ref{1aug2})-(\ref{1aug4}) --- this already ensures that the dynamical field $\Phi$ carries helicity $\lambda$, while the remaining fields are either auxiliary or pure gauge. Then we can proceed differently. It seems that the most natural way to achieve correct transformation properties is to declare that ``$i$'' of $A_i$ is a Lorentz form index. This already ``contributes'' helicity one to  transformation properties of $A^x$. It then remains to declare that $A_i$ is a 1-form that is additionally valued in the representation of the Lorentz group given by (\ref{1aug2})-(\ref{1aug4}), but with $\lambda$ replaced by $\lambda-1$.

This approach is to some extent reminiscent of the frame-like approach \cite{Vasiliev:1980as,Fradkin:1987ks,Fradkin:1986qy},
where higher spin fields are also represented by one-forms taking values in representations of the Lorentz algebra. The difference, however, is that while in the frame-like approach these representations are realised by multi-component tensors, in the reformulation that we found here, the fibre space is given by reals, which, nevertheless, transform as helicity $\lambda-1$ light-cone fields.
For this reason,  formulation (\ref{31aug18})
can be characterised as {\em{semilight-cone approach}} --- 
 light-cone gauge is imposed in the fibre, but not on the base.

On the practical side, a major advantage of the semilight-cone gauge over the usual frame-like approach is that it allows
to deal with all vertices, including those, which are absent in the manifestly covariant approach. Another 
advantage is that, as it follows from the above discussion,  it allows 
a completely uniform treatment of all massless theories at least in 
the self-dual sector. In contrast, in the frame-like approach this can be achieved only in AdS and using some 
supplementary constructions \cite{MacDowell:1977jt,Stelle:1979aj}. Finally, we note that connections $A_i$ seem to be closely
related to the fields employed in the twistor construction of the conformal higher spin theory \cite{Haehnel:2016mlb,Adamo:2016ple}.

\subsection{Hidden symmetries from 2d sigma models}
\label{section6.1}
The actual self-dual Yang-Mills equations are known to be integrable. As a result they have a list of remarkable properties.
In particular, they posses an infinite set of conserved quantities associated with an infinite hidden symmetry algebra.
It is also possible to construct exact solutions of these equations via certain solution generating techniques and these 
solutions can often be superposed. Understanding properties of the self-dual Yang-Mills equations was one of the original
motivations for the twistor geometric construction  \cite{Mason:1991rf,Hitchin:1999at}.

It is natural to expect that these features in some form extend to other chiral cubic  theories.
In the following we will focus on one particular aspect of equations (\ref{31aug18}):  we will derive their hidden 
symmetry algebras. To do that, following Park \cite{Park:1989vq,Park:1990fp}, we rewrite them as equations of motion for 
some 2d sigma model and then use known results for hidden symmetries of this sigma model. This approach
allows a uniform treatment of all chiral cubic theories as well as reveals their two-dimensional nature.
 But first, let us review the results for lower spin cases.

The quest for hidden symmetries of the self-dual Yang-Mills equations developed in the following way. First, it was realised
that the construction of an infinite set of conserved currents in 2d sigma models \cite{Luscher:1977rq,Brezin:1979am}
can be straightforwardly generalised to the self-dual Yang-Mills equations \cite{Prasad:1979zc,Pohlmeyer:1979ya}.
Then the associated symmetry algebra was found in \cite{Chau:1981gi,Chau:1982mj,Dolan:1982dc}.
The problem of hidden symmetries of the self-dual Yang-Mills equations was also addressed using 
its reformulation as  the Riemann-Hilbert problem
\cite{Ueno:1982dy}
and using the twistor geometry \cite{Crane:1987im}.
Eventually, it was found that for the self-dual Yang-Mills equations with the internal symmetry algebra $\mathfrak{g}$ 
the hidden symmetry algebra is
\begin{equation}
\label{31aug20}
\mathfrak{F}_{\rm YM} = \mathfrak{g}\otimes C(x^--\zeta \bar x,x+\zeta x^+,\zeta,\zeta^{-1}),
\end{equation}
where $\zeta\in \mathbb{C}$ is an auxiliary variable.

The algebra of hidden symmetries of the self-dual Einstein equations was first found in \cite{Boyer:1977,Boyer:1985aj,Boyer:1989}
based on the reformulation of these equations  by Plebanski \cite{Plebanski:1975wn}
\begin{equation}
\label{31aug21}
\mathfrak{F}_{\rm GR} = {\rm SDiff}(x^-,x)\otimes C(\zeta,\zeta^{-1}).
\end{equation}
This result was later reproduced by various methods, see  \cite{Takasaki:1989cg,Park:1989vq,Park:1990fp,Grant:1992ba,Husain:1993dp,Husain:1994dw,Husain:1994vi,Strachan:1994ts}. 

The approach that we found particularly convenient for the analysis of (\ref{31aug18}) is to 
 present these equations first in the form of field equations for the 2d sigma model with the pure  Wess-Zumino term.
For the self-dual Einstein equations this was done in  \cite{Park:1989vq} and then generalised to include the self-dual
Yang-Mills equations in \cite{Park:1990fp}. Similarly, in \cite{Husain:1993dp,Husain:1994dw,Husain:1994vi} the self-dual
Einstein equations were rewritten in the form of equations of motion for the 2d principal chiral sigma model.

Let us remind the reader that equations of motion for the principal chiral sigma model have the form
\begin{equation}
\label{3sepsm1}
\partial_0 \big(g^{-1}\partial_0 g \big) + \partial_1 \big( g^{-1} \partial_1 g\big)=0
\end{equation}
and those produced by the Wess-Zumino term are
\begin{equation}
\label{3sepsm2}
\partial_0 \big(g^{-1}\partial_1 g \big) - \partial_1 \big( g^{-1} \partial_0 g\big)=0 . 
\end{equation}
Here $g$ is a group element of some internal Lie group.

Once such a reformulation is achieved, one can use the known results for hidden symmetries of
 sigma models. In particular, the hidden symmetry algebra of the principal chiral sigma model
 with an internal algebra $\mathfrak{g}$ is given by the loop extension of the latter
 \cite{Luscher:1977rq,Brezin:1979am,Dolan:1981fq,Devchand:1981wy,Ueno:1982av,Wu:1982jt}
 \begin{equation}
 \label{3semsm3}
 \mathfrak{F} = \mathfrak{g}\otimes C(\zeta,\zeta^{-1}).
 \end{equation}
 At the same time, the pure Wess-Zumino term sigma model can be related to the principal one
 by a certain non-local transformation and their hidden symmetry algebras coincide, see \cite{Schwarz:1995td}.
 We also refer the reader to \cite{Schwarz:1995td} for a pedagogical review on the subject.
 
To rewrite (\ref{31aug18}) in the desired form we start by fixing the gauge $A^-=0$. 
The first  equation of (\ref{31aug18}) then entails $A^x=0$. The other 
two equations become
\begin{align}
\notag
\partial^- A^+ + \partial^x A^{\bar x}=0,\\
\partial^+ A^{\bar x}-\partial^{\bar x} A^+ + g[A^+, A^{\bar x}]=0.
\label{3sep1}
\end{align}
Then we replace fields by operators that result
from their adjoint action in respective algebras
\begin{equation}
\label{3sep6}
A(x;z) \quad \to \quad \hat A(x;z) \equiv [A(x,z), \cdot].
\end{equation}
Employing the Jacobi identity, we can rewrite (\ref{3sep1}) identically just adding hats to fields
\begin{align}
\notag
\partial^- \hat A^+ + \partial^x \hat A^{\bar x}=0,\\
\partial^+ \hat A^{\bar x}-\partial^{\bar x} \hat A^+ + g[\hat A^+, \hat A^{\bar x}]=0.
\label{3sep06}
\end{align}

 Next, we introduce operators
\begin{equation}
\label{3sep7}
\hat K^- \equiv - \partial^{\bar x} - g\hat A^{\bar x}, \qquad K^x \equiv \partial^+ + gA^+.
\end{equation}
In these terms (\ref{3sep06}) reads
\begin{align}
\notag
\partial^- \hat K^x - \partial^x \hat K^- =0,\\
\label{3sep8}
[\hat K^-,\hat K^x]=0,
\end{align}
which imply 
\begin{equation}
\label{3sep9}
\partial^- \hat K^x - \partial^x \hat K^-+ g [\hat K^-,\hat K^x]=0.
\end{equation}

Now we would like to interpret $\bar x$ and $x^+$ as  base coordinates and $K_{\bar x}$, 
$K_{+}$ as the associated components of the connection. The Lie algebra where this connection
takes values is defined by the commutator of the chiral cubic  theory that we 
started from, followed by
 redefinition (\ref{3sep7}). The elements of this algebra are given by differential operators
involving only derivatives  $\partial_-$ and $\partial_x$. Therefore, we can interpret $x^-$
and $ x$ as fibre coordinates and the connection acts by differential operators on the space parametrised
by these two coordinates.
In other words, we managed to achieve an explicit splitting of base coordinates $\bar x$ and $x^+$
and fibre coordinates $x^-$ and $x$.

With this clarified, we can interpret (\ref{3sep9}) as a zero curvature condition for the connection $K$, 
which implies
\begin{equation}
\label{3sep10}
K_\mu = g^{-1}\partial_\mu g,
\end{equation} 
where $\mu$ takes two values $x^{\mu} = \{x^+,\bar x \}$
 and $g$ is a group element associated with the internal Lie algebra. Plugging (\ref{3sep10})
 into the first equation of (\ref{3sep8}) we find equations of motion for the sigma model with the 
 pure Wess-Zumino term, see (\ref{3sepsm2})
 \begin{equation}
 \label{3sep11}
 \partial^- \big(g^{-1}\partial g \big) - \partial \big( g^{-1} \partial^- g\big)=0.
 \end{equation}

 This entails the following hidden symmetries for chiral higher spin theories. The chiral higher spin theory
 and the Poisson chiral higher spin theory both have graviton in the spectrum. Then (\ref{3sep7})
 can be regarded as a redefinition of graviton and, hence, the algebra generated by $K$ is the same
 as the algebra generated by $A$. In other words, the hidden symmetry algebra of 
 the chiral higher spin theory is, see (\ref{3semsm3}),
 \begin{equation}
 \label{3sep12}
 \mathfrak{F}_{\rm HS} = {\rm hs}(x^-,x)\otimes C[\zeta,\zeta^{-1}]
 \end{equation}
 and of the Poisson chiral higher spin theory is
 \begin{equation}
 \label{3sep13}
 \mathfrak{F}_{\rm PHS} = {\rm phs}(x^-,x)\otimes C[\zeta,\zeta^{-1}].
 \end{equation}
 
 On the contrary, for the coloured higher spin theory graviton is not in the spectrum and $K$ do not form
 the same algebra as $A$. Still the shift (\ref{3sep7}) can be easily taken into account analogously
 to the case of the self-dual Yang-Mills theory (\ref{31aug20}). The resulting  hidden symmetry algebra is 
 \begin{equation}
 \label{3sep14}
 \mathfrak{F}_{\rm CHS} = {\rm chs}(x^--\zeta \bar x,x+\zeta x^+,\zeta,\zeta^{-1})
 =\mathfrak{g}\otimes C(z,z^{-1})\otimes C(x^--\zeta \bar x,x+\zeta x^+,\zeta,\zeta^{-1}).
 \end{equation}

Before closing the section, let us give another form of equations (\ref{31aug18}),
that generalises the Plebanski equations \cite{Plebanski:1975wn} to higher spins.
Now we solve the first equation in (\ref{3sep1}) as
\begin{equation}
\label{3sep2}
A^+ = \partial^x \omega, \qquad A^{\bar x}= -\partial^- \omega.
\end{equation}
Plugging this into the second equation we find
\begin{equation}
\label{3sep3}
(\partial^+\partial^-+\partial^{\bar x}\partial^x)\omega + g [\partial^x \omega, \partial^- \omega]=0.
\end{equation}
Then, we observe that (\ref{3sep3}) can be rewritten as
\begin{equation}
\label{3sep4}
[\partial^x \Omega,\partial^- \Omega]=z,
\end{equation}
where
\begin{equation}
\label{3sep5}
\Omega = \Omega_0 + g\omega, \qquad \Omega_0 = z(x\bar x+ x^+ x^-).
\end{equation}
Here $\Omega_0$ should be interpreted as a background field. We can see from the degree of $z$ 
in (\ref{3sep5}) that only helicity two field, that is graviton, has a non-zero background value. 
Equation (\ref{3sep4}) can be regarded as a higher spin generalisation of first  Plebanski's  heavenly 
equation \cite{Plebanski:1975wn}. 

It is also worth to note that in many manipulations in this section we were sloppy with counting functions
of two variables. For example, when we fix  light-cone gauge $A^+=A^{\bar x}=0$ and do not
impose any boundary conditions, there is a residual gauge symmetry with a parameter depending
on $x^+$ and $\bar x$. Similar issue occurs when we solve the first of equations (\ref{3sep1})
as (\ref{3sep2}). As another example, operation (\ref{3sep6}) of replacement of $A$ by the associated operator has
a non-zero kernel consisting of functions of $x^+$ and $\bar x$.
These and similar issues are more carefully discussed, for instance, in \cite{Mason:1991rf,Hitchin:1999at,Park:1989vq,Park:1990fp}.
We would like to note, however, that in 4d each on-shell degree of freedom is characterised by a function of three
variables (for example, one can choose $q$, $\bar q$ and $q^+$), so from the point of view of controlling the degrees of freedom, our analysis is completely accurate.

\section{Colour-kinematics duality}
\label{colkin}
One important context where  the algebra $\mathfrak{G}_{\rm GR}$  defined above previously appeared  
 is  the colour-kinematics duality in the self-dual sector, where $\mathfrak{G}_{\rm GR}$
 was identified as the gauge theory kinematic algebra and was also connected  to 
 hidden symmetries of the self-dual gravity equations \cite{Monteiro:2011pc}.
Hidden symmetries in self-dual theories have been already analysed in the previous section.  Here we focus on the colour-kinematics 
 duality.

The colour-kinematics duality  implies a set of relations for the Yang-Mills amplitudes \cite{Bern:2008qj}, called the BCJ relations,
that allow to substantially reduce the number of independent colour ordered amplitudes. It also gives a simple squaring
procedure, that allows to generate gravity amplitudes out of amplitudes of gauge theory \cite{Bern:2008qj,Bern:2010ue,Bern:2010yg}.
Below we give a brief review of how the colour-kinematics duality works. 

To start, one has to reorganise a gauge theory amplitude ${\cal A}_n$ in a way that it appears as a 
sum over cubic diagrams, that is diagrams involving only cubic vertices
\begin{equation}
\label{29aug13}
{\cal A}_n = g^{n-2}\sum_{j} \frac{c_j n_j}{D_j}.
\end{equation}
Here $j$ enumerates all possible types (channels) of cubic diagrams, $D_j^{-1}$ is the product
of propagators associated with diagram type $j$, $c_j$ contains all amplitude's colour dependence, while $n_j$ 
contains the remaining dependence on kinematical data. Factors $c_j$ and $n_j$ are called 
\emph{colour} and \emph{kinematic numerators} respectively. To achieve representation (\ref{29aug13})
one has to split all quartic vertices that appear in the standard Feynman diagram expansion accordingly
to their colour structure into four-point exchanges. This is done simply by dividing and then multiplying the 
quartic vertex by missing propagators.

Due to quartic vertices splitting ambiguities  the resulting representation in terms of cubic diagrams
 is not unique. It turns out that this ambiguity can be fixed in such a way that whenever we pick three diagrams 
with their colour factors satisfying the generalised Jacobi identity
\begin{equation}
\label{30aug1}
c_{j_1}+c_{j_2}+c_{j_3}=0,
\end{equation}
the associated kinematic factors also satisfy the Jacobi identity
\begin{equation}
\label{30aug2}
n_{j_1}+n_{j_2}+n_{j_3}=0.
\end{equation}
For the simplest case of four-point amplitudes, $j_1$, $j_2$ and $j_3$ are just the $s$-, $t$- and $u$-channel
cubic diagrams, while (\ref{30aug1}) is the usual Jacobi identity for the internal symmetry algebra. At higher points
(\ref{30aug1}) is satisfied when cubic diagrams 
$j_1$, $j_2$ and $j_3$
 differ only by four-point subdiagrams, which are, in turn, given by three different cubic diagrams connecting these
 four points.
  Then the Jacobi identity (\ref{30aug1})
is satisfied as a consequence of the usual Jacobi identity for the four-point subdiagrams.

One consequence of relations (\ref{30aug2}) is that they allow to reduce the basis of Yang-Mills partial amplitudes.
They also point towards existence of the so-called kinematic algebra associated with the Jacobi identity (\ref{30aug2}),
which is expected to govern the structure of  Yang-Mills amplitudes similarly to how  it does the colour Lie algebra, 
associated with the Jacobi identity (\ref{30aug1}).
Moreover, 
once the cubic representation (\ref{29aug13}) for a gauge theory amplitude is achieved
and the BCJ relations (\ref{30aug2}) are satisfied, one can obtain
the associated gravity amplitude by a simple squaring procedure, which amounts to replacing colour 
factors $c_j$ by their kinematic counterparts $n_j$
\begin{equation}
\label{30aug3}
{\cal M}_n = g^{n-2}\sum_{j} \frac{n_j n_j}{D_j}.
\end{equation}
This double copy construction turns out be an efficient tool for generating (super)-gravity amplitudes from
(super)-Yang-Mills ones. Quite remarkably, it also works at the loop level
 and has many other applications in quite a broad
range of theories, for review see \cite{Elvang:2013cua}.
Still the meaning of the kinematic algebra in these cases remains unclear.

The situation is much more transparent for the self-dual Yang-Mills equations, 
for which the kinematic algebra was identified in \cite{Monteiro:2011pc}. What makes self-dual theories
simpler is  that their  Lagrangians  (\ref{8aug1}), (\ref{24aug2}) contain only cubic vertices from the very beginning.
 This implies that the Feynman rules immediately
produce amplitudes in the cubic representation. This, in turn, means that the kinematic structure constants can be
read off straight from the self-dual Yang-Mills  action. More precisely,  to find out 
kinematic structure constants one just needs to factor out colour factors from the self-dual Yang-Mills cubic vertices.
This results in 
\begin{equation}
\label{30aug4} 
f^{A_1}{}_{A_2A_3} =  \delta(-q_1+q_2+q_3) \frac{\bar{\mathbb{P}}_{23}}{\beta_2\beta_3}\beta_1 \delta^{\lambda_1,2}\delta_{\lambda_2,2}\delta_{\lambda_3,2}.
\end{equation}
These are exactly the structure constants that we previously associated with self-dual gravity (\ref{24aug3}) and they define
the algebra $\mathfrak{G}_{\rm GR}$. Moreover, by squaring the kinematic numerators
 of the self-dual Yang-Mills theory one finds the numerators for the associated 
gravity amplitudes as a consequence of this relation being true for cubic vertices. Before closing this review part,
  it is worth to emphasise that all amplitudes
in the self-dual theories vanish on-shell, so to make this discussion non-trivial one should consider amplitudes with
off-shell momenta on external lines.

\subsection{Generalisation to higher spin theories}
\label{ckhs}
In the context of the colour-kinematics duality one way to interpret our results on the structure of cubic vertices
in chiral theories is as follows.
We can still split cubic vertices of chiral theories as we did in section \ref{sect3}. Then for a given cubic diagram,
the kinematic numerator of a chiral theory can be defined to be equal to the kinematic numerator
 of the self-dual Yang-Mills theory computed for the
same cubic diagram. In turn, the colour numerator can be defined as a product of the gauge algebra structure constants 
over all vertices in this cubic diagram. Then both (\ref{30aug1}) and (\ref{30aug2}) are automatically satisfied. Indeed, (\ref{30aug1})
is satisfied because, as we checked, the gauge algebra structure constants satisfy the Jacobi identity, while
(\ref{30aug2}) is satisfied because of an accidental equality between the kinematic part of the Yang-Mills vertex and the
structure constants of  self-dual gravity. Thus, amplitudes of chiral cubic theories indeed satisfy relations that can be 
regarded as the generalised BCJ relations.

It is also interesting to see how the generalised squaring procedure that would include higher spin theories might work\footnote{At the thee-point level generalisation of the colour-kinematics duality to higher spins
 was previously discussed in \cite{Ananth:2012un,Akshay:2014qea}.}.
Of course, the most attractive scenario would be if this procedure  allowed  to compute higher spin amplitudes 
from lower spin ones. In Appendix \ref{gsp} we show that it is, indeed, possible to generate amplitudes of chiral higher 
spin theories by multiplying self-dual Yang-Mills amplitudes. However, this construction goes too far away
from how the original colour-kinematics duality works, so it does not seem very likely that it can be extended to parity invariant
theories.

Instead, it seems that the following version of the higher spin colour-kinematics duality is more realistic.
We recall that the BCJ relations follow from properties of  underlying open string theory amplitudes
\cite{BjerrumBohr:2009rd,Stieberger:2009hq}, while the double copy construction is proven using the relation between open 
and closed string theory amplitudes \cite{Kawai:1985xq}. From its general structure, it seems natural to assume that a putative parity invariant
completion of the coloured chiral higher spin theory, similarly to the Yang-Mills theory, may have an underlying open string
description. Then, coloured higher spin amplitudes should satisfy the BCJ relations with the colour and 
kinematic numerators defined in the standard way: colour numerators consist of products of all internal symmetry
colour factors, while kinematic numerators accumulate all  kinematic dependence contributed by vertices. For example, for the
coloured chiral higher spin theory these are of the form, see (\ref{28aug2})
\begin{equation}
\label{17sep1}
c_{j} \equiv \prod_V f^{\alpha_1\alpha_2\alpha_3}, \qquad n_j \equiv \prod_V \delta(-q_1+q_2+q_3)\frac{ \delta (-\check\lambda_1+\check\lambda_2+\check\lambda_3)}{(-\beta_1)^{-\check\lambda_1}
\beta_2^{\check \lambda_2}\beta_3^{\check\lambda_3}}\bar{\mathbb{P}}_{23},
\end{equation}
where the product is over all vertices in the diagram. Then, by the double copy construction we expect to produce amplitudes of
the putative parity invariant completion of the Poisson chiral higher spin theory.

Of course, if the higher spin colour-kinematics duality works in this form, the resulting constraints are not yet sufficient 
to define any of the two theories. Still, for the coloured higher spin theory, similarly to  Yang-Mills theory, 
they constrain the number of independent partial amplitudes
to $(n-3)!$ at $n$ points, which may be helpful. Possibly, combined with other considerations, the
BCJ relations will allow to generate higher 
order higher spin  amplitudes without really knowing the complete Lagrangian.

Finally, let us note another intriguing possibility that may allow to compute higher spin amplitudes without
knowing the  Lagrangian.
In \cite{Chalmers:1998jb} Chalmers and Siegel proposed 
to replace light-cone gauge by another axial gauge with the axial direction defined by momenta of scattered
particles in some special way. The advantage of this approach is that at least for
 Yang-Mills theory and gravity, contributions from quartic vertices 
vanish. It would be interesting to see whether this idea can be consistently generalised to include higher spin theories 
and to compute higher point amplitudes knowing only cubic vertices.
The approach of Chalmers and Siegel was already used fruitfully to prove colour-kinematics duality for MHV amplitudes in  
Yang-Mills theory and gravity \cite{Monteiro:2011pc}.

\section{Universality}
\label{universality}
In section \ref{sect3} we considered many examples of chiral  cubic  theories and demonstrated that the same pattern
systematically takes place: the vertices
have the form of a product of the kinematic Yang-Mills vertex and the structure constants, the latter satisfying the Jacobi
identity.  So far it was a case by case analysis and the reason why this pattern reappears each time was not clear.
In this section we give a universal argument showing that the light-cone consistency conditions imply the Jacobi identity
for the gauge algebra structure constants.\footnote{A connection between consistency of the four-particle scattering with
the Jacobi identity in Yang-Mills theory was also found in other approaches \cite{Benincasa:2007xk,McGady:2013sga,Ananth:2017pio}.
We would like to emphasise that these approaches where based on the analysis of amplitudes featuring one holomorphic
and one antiholomorphic vertex. For higher spin theories such an analysis cannot be applied as it faces a locality obstruction. Still, 
as we demonstrate, the gauge algebra  can be recovered already from consistency of the self-dual sector.}

The light-cone consistency condition (\ref{1aug18}) can be rewritten as 
\begin{equation}
\label{4sep1}
[{\cal A}, J_2]=0,
\end{equation}
where $J_2$ is the quadratic part of the charge $J$ and ${\cal A}$ is the off-shell amplitude constructed from the
light-cone Hamiltonian in a specific way.\footnote{For more details, see  \cite{Ponomarev:2016cwi}. See also
 \cite{Metsaev:1991mt} for a similar statement for four-point functions.}
 For  chiral cubic theories it is enough to check the consistency of the four-point amplitude, in which 
 case (\ref{4sep1}) reads
\begin{equation}
\label{4sep2}
\delta(\sum q^\perp_i)
{\cal J}\left( \sum_{\lambda_e}h_3^{\lambda_1,\lambda_2,-\lambda_e}(q_1,q_2,Q_{12,34})\frac{1}{s_{12,34}}h_3^{\lambda_3,\lambda_4,\lambda_e}(q_3,q_4,Q_{34,12})+
\text{2 terms}  \right)=0,
\end{equation}
where 
\begin{equation}
\label{4sep3}
Q_{ij,kl}=\frac{1}{2}(-q_i-q_j+q_k+q_l), \qquad s_{ij,kl}=\frac{1}{2}([ij]\langle ij \rangle + [kl]\langle kl \rangle), 
\end{equation}
$\lambda_e$ is helicity of the exchanged particle and ${\cal J}$ was given in (\ref{1aug21}). Spinor-helicity
conventions are given in Appendix \ref{conventions}.

In the following we will use notation ${\cal J}^{\lambda_1,\dots,\lambda_4}$ to make dependence of ${\cal J}$ on helicities explicit.
Employing the chain rule we easily find that
\begin{equation}
\label{4sep4}
{\cal J}^{\lambda_1+\lambda_1',\lambda_2+\lambda_2',\lambda_3+\lambda_3',\lambda_4+\lambda_4'}(A \cdot B) = 
{\cal J}^{\lambda_1,\lambda_2,\lambda_3,\lambda_4}A \cdot B +
A\cdot {\cal J}^{\lambda'_1,\lambda'_2,\lambda'_3,\lambda'_4}B
\end{equation}
for any $\lambda$ and $\lambda'$ as well as for any four-point functions $A$ and $B$. We will also need that \cite{Ponomarev:2016cwi}
\begin{equation}
\label{4sep5}
[{\cal J}^{0,0,0,0}, s_{ij,kl}]=0.
\end{equation}

To proceed, we factorise vertices  into a product of the kinematic part of the Yang-Mills vertex and the structure constants, 
as defined in (\ref{6aug7})
\begin{align}
\notag
h_{3}^{\lambda_1,\lambda_2,-\lambda_e}(q_1,q_2,Q_{12,34})\; & = f^{\lambda_1,\lambda_2,-\lambda_e}(q_1,q_2,Q_{12,34})h_{{\rm YM}}^{1,1,-1}(q_1,q_2,Q_{12,34})\\
\label{4sep6}
h_{3}^{\lambda_3,\lambda_4,\lambda_e}(q_3,q_4,Q_{34,12}) \; &= f^{\lambda_3,\lambda_4,\lambda_e}(q_3,q_4,Q_{34,12}) h^{1,-1,1}_{\rm YM}(q_3,q_4,Q_{34,12})
\end{align}
and similarly for other channels.
In agreement with our condensed notations we will denote
\begin{equation}
\label{4sep7}
f_{A_1A_2}{}^{E}\equiv f^{\lambda_1,\lambda_2,-\lambda_e}(q_1,q_2,Q_{12,34}), \qquad
f_{A_3A_4 E} \equiv f^{\lambda_3,\lambda_4,\lambda_e}(q_3,q_4,Q_{34,12}),
\end{equation}
where
\begin{equation}
\label{4sep8}
 E \equiv \{\lambda_E, -Q_{12,34} \}.
\end{equation}

In these terms, employing (\ref{4sep4}), (\ref{4sep5}), the consistency condition (\ref{4sep2}) can be rewritten as
\begin{align}
\notag
\delta(\sum q^\perp_i)\left(\vphantom{\frac{1}{s_{12,34}} }\right.  \Big[ h_{{\rm YM}}^{1,1,-1}(q_1,q_2,Q_{12,34})\frac{1}{s_{12,34}}   h^{1,-1,1}_{\rm YM}(q_3,&\: q_4,Q_{34,12})\Big] \\
   \cdot
\notag
{\cal J}^{\lambda_1-1,\lambda_2-1,\lambda_3-1,\lambda_4+1} &\left[ f_{A_1A_2}{}^E f_{A_3A_4E}\right]\\
\notag
+{\cal J}^{1,1,1,-1} \Big[  h_{{\rm YM}}^{1,1,-1}(q_1,q_2,Q_{12,34})\frac{1}{s_{12,34}}   h^{1,-1,1}_{\rm YM}(q_3,&\:q_4,Q_{34,12})\Big] \\
\label{4sep9}
 \cdot & \left[ f_{A_1A_2}{}^E f_{A_3A_4E}\right]
+  \left.      \vphantom{\frac{1}{s_{12,34}} }   \text{2 terms}  \right)=0.
\end{align}
Schematically, it is of the form
\begin{equation}
\label{4sep10}
{\cal O}_s \left[ f_{A_1A_2}{}^E f_{A_3A_4E}\right] + {\cal O}_t \left[ f_{A_2A_3}{}^E f_{A_1A_4E}\right] +
{\cal O}_u \left[ f_{A_3A_1}{}^E f_{A_2A_4E}\right]=0. 
\end{equation}
If we manage to show that 
\begin{equation}
\label{4sep11}
{\cal O}\equiv {\cal O}_s={\cal O}_t = {\cal O}_u
\end{equation}
then, clearly, the light-cone consistency condition in the form (\ref{4sep10}) implies the Jacobi identity up to terms in the kernel of ${\cal O}$.

To show (\ref{4sep11}), first, we notice that 
\begin{equation}
\label{4sep12}
{\cal J}^{1,1,1,-1}\Big[  h_{{\rm YM}}^{1,1,-1}(q_1,q_2,Q_{12,34})\frac{1}{s_{12,34}}   h^{1,-1,1}_{\rm YM}(q_3,q_4,Q_{34,12})\Big]
\end{equation}
is, indeed, channel-independent as it is exactly the consistency condition (\ref{4sep9}) for the self-dual Yang-Mills theory.
On the other hand, 
\begin{equation}
\label{4sep13}
 h_{{\rm YM}}^{1,1,-1}(q_1,q_2,Q_{12,34})\frac{1}{s_{12,34}}   h^{1,-1,1}_{\rm YM}(q_3,q_4,Q_{34,12})
\end{equation}
when compared between different channels differs by terms proportional to ${\cal H}$, see (\ref{1aug21}) for definition.
This is a consequence of the fact that the four-point amplitude of the self-dual Yang-Mills theory vanishes on-shell and can also
be easily checked explicitly. 

In other words, we managed to show that ${\cal O}_s$ is equal to ${\cal O}_t$ and ${\cal O}_u$ up to terms proportional to ${\cal H}$.
These terms, in turn, can be compensated by using momentum conservation inside the Jacobi identity. Indeed, one can see that
\begin{equation}
\label{4sep14}
\delta(\sum q^\perp_i) [{\cal J}, \alpha(q^\perp) \sum(\bar q_i)] = \delta(\sum q^\perp_i) \alpha(q^\perp){\cal H}
\end{equation}
for any function $\alpha(q^\perp)$ of transverse momenta. Therefore, by adding total derivatives to a product of structure constants in 
the second line of (\ref{4sep9}) and similarly for other channels, one can compensate all mismatches between ${\cal O}_s$, ${\cal O}_t$ and ${\cal O}_u$.

To summarise, we managed to show that the light-cone consistency condition can be written in the form (\ref{4sep10}), (\ref{4sep11}).
To achieve this, one should  properly employ momentum conservation inside the Jacobi identity. 
In turn, (\ref{4sep10}), (\ref{4sep11})  implies that the Jacobi identity holds
up to terms in the kernel of ${\cal O}$. This is what we intended to 
show.

A couple of comments are now in order. It is not hard to show that ${\cal O}{\cal H}\propto {\cal H}$, that
is contributions to the Jacobi identity proportional to ${\cal H}$ give rise to terms proportional to ${\cal H}$ in (\ref{4sep10}).
 These contributions can be compensated by
using momentum conservation inside the Jacobi identity as done above. This, in turn, implies that terms proportional to ${\cal H}$
are, effectively, in the kernel of ${\cal O}$ and, hence, the Jacobi identity can be proven only up to such terms. Another way to 
put this is that the argument above proves the Jacobi identity  only for on-shell external momenta.
At the same time, from numerous examples we considered in section \ref{sect3} one can see that the Jacobi identity for the gauge algebra
also holds off-shell. This suggests  that it should be possible to extend the proof given above to off-shell momenta.

Another interesting direction to extend the above analysis is to include higher vertices. From comparison with the covariant approach
we can anticipate that the underlying algebraic structure responsible for symmetries of chiral theories with higher vertices is
 ${\rm L}_\infty$ algebra\footnote{For some previous discussions on the connection between  ${\rm L}_\infty$ and the 
 light-cone deformation procedure see \cite{Bengtsson:2004cd,Bengtsson:2006pw}.},
  see e.g. \cite{Lada:1992wc} for review. Similarly to cubic theories, we
 expect that Lorentz invariance implies a system of  ${\rm L}_\infty$ relations for the associated structure constants.
Moreover, if a simple and unambiguous relation between vertices and structure constants extends from cubic order 
to higher orders one can, in principle,
facilitate the analysis of the light-cone consistency conditions by replacing them with the analysis of the corresponding 
${\rm L}_\infty$ relations.

\section{Conclusion}
\label{conclusion}
In this paper we studied chiral higher spin theories. We demonstrated that they can 
be regarded as particular members of a class of theories that we call {\em{chiral cubic  theories}}. 
All these theories contain only cubic vertices and only of one type --- either made of angle or of square spinor brackets.
The structure of the consistency conditions in the light-cone deformation procedure implies that these theories
 do not require any completion with higher degree vertices.
We show that all chiral cubic  theories can be regarded as generalised Yang-Mills theories
with some gauge algebra, which may involve space-time derivatives.

In the context of the light-cone deformation procedure where the gauge freedom is completely fixed, 
the notion of a gauge algebra deserves clarification. One way that we can define it is directly from the light-cone
 Lagrangian: the gauge algebra structure constants can be obtained by factoring out the kinematic part of the
 Yang-Mills vertex from the vertex of a given chiral cubic  theory. Alternatively, one can partially undo the light-cone
 gauge fixing and demonstrate that the equations of motion in these theories can be written as the self-duality
 conditions for the Yang-Mills curvature associated with a gauge algebra in a standard way. 
 
 It is worth to stress 
 that the procedure of undoing  light-cone gauge that is required to reveal the self-dual nature of 
 chiral cubic  theories treats fields as if they were all the Yang-Mills 
 connections rather than higher spin fields. In particular, the set of fields that we eventually obtain for a helicity $\lambda$ field
  is a four-component
 vector, valued in helicity-$(\lambda-1)$ representation of the Lorentz group, rather than a double-traceless rank $\lambda$ symmetric tensor of Fronsdal's approach.
 We find this new field content for higher spin fields rather attractive, because, while, on the one hand, it leaves room
 for gauge symmetries and related geometric constructions such as the Yang-Mills curvature, on the other hand, 
it still allows to construct all possible cubic interactions without violating locality. In other words, this approach
seems to have benefits of both the covariant  and the light-cone approaches.
  It would be interesting to see if 
the aforementioned  curvatures can be useful to construct  actions for parity invariant theories. This idea has some common features
and is, possibly, related to the approach of quadratic forms   \cite{Ananth:2015tsa,Ananth:2017xpj}.

Among chiral cubic  theories we made a particular focus on three higher spin theories.
Two of them --- the chiral higher spin theory and its coloured version --- were proposed in  \cite{Ponomarev:2016lrm}
based on the earlier analysis  \cite{Metsaev:1991mt,Metsaev:1991nb,Metsaev:1991thesis}. We discussed the associated
gauge algebras and pointed out some novel properties that distinguish them from their AdS counterparts. The third theory
was found in this paper and can be regarded as a result of the Poisson contraction of the chiral higher spin theory. Its Lagrangian
contains only two-derivative terms, which makes it reminiscent of gravity. We also point out similarities between gauge algebras we
found here and those found in AdS by solving the Jacobi identity \cite{Boulanger:2013zza}.

Next, we studied what the connection of chiral cubic  theories with the self-dual Yang-Mills equations
implies for these theories.
The self-dual Yang-Mills equations are known to posses an infinite  hidden symmetry  algebra, which also
implies integrability of these equations. Using further reformulation of chiral cubic theories as certain 2d
sigma models we found their hidden symmetry algebras. Presence of an infinite symmetry typically implies
triviality of the associated S-matrix. At least, this is the case for the self-dual Yang-Mills theory and gravity. The question whether this
is also true for chiral higher spin theories we leave for future research.

 Reformulation of equations of motion of chiral higher spin theories as the self-dual Yang-Mills equations also 
  allows to interpret them as consistency conditions for the associated linear problems, which, in turn,
serves as a starting point for various solution generating techniques \cite{Mason:1991rf,Hitchin:1999at}.
We leave the interesting question of higher spin solitons for future research. Another important question that  would
be interesting to clarify in future is the twistor geometry that underlies chiral higher spin theories. 
In this respect, let us note
that recently conformal higher spin theory was constructed and further studied using twistor space 
techniques  \cite{Haehnel:2016mlb,Adamo:2016ple}.

Another interesting context to which we connect our observations is the colour-ki\-ne\-ma\-tics duality
 \cite{Bern:2008qj,Bern:2010ue,Bern:2010yg}. 
This duality takes an especially simple form  for self-dual theories \cite{Monteiro:2011pc}. 
 We show that  in the self-dual sector the colour-kinematics duality can be
naturally generalised to include chiral higher spin theories. In particular, we propose generalised 
double copy procedures that relate chiral higher spin theories to each other as well as to the self-dual
Yang-Mills theory.  We leave an interesting question of extension
of these results to parity invariant theories for future research.

We also explain why the universal relation between chiral cubic  theories and the self-dual
Yang-Mills theory takes place. More precisely, we prove that the Jacobi identity for the gauge 
algebra in cubic chiral theories holds as a consequence of the light-cone consistency conditions or, in other words, of Lorentz
invariance. We find instructive to view this as an analogue of a relation between gauge invariance of the action 
and closure of the Jacobi identity in the covariant approach. However, the mechanisms of how this connection appears in
two approaches is completely different. Contrary to the covariant approach, in the light-cone deformation procedure the gauge
freedom is completely fixed already at free level, so presence of any symmetry algebras for light-cone theories is much more implicit.
One direction where this result can be extended is to include higher order vertices. Similarly to covariant approaches,
we expect the underlying algebraic structure to generalise to ${\rm L}_\infty$ algebras.

Finally, we make some comments on (non)-locality in parity invariant theories. By now a lot of evidence is accumulated to conclude that
irrespectively of the approach, one cannot avoid non-localities for parity invariant higher spin theories. In particular, in 
Appendix \ref{locobstr} we review the results of \cite{PonSkv:2016}, where a local obstruction to parity invariant minimal gravitational interactions
of higher spin fields was found using the light-cone deformation procedure.
Non-localities for a good reason have bad reputation in physics. They typically lead to undesirable consequences, which 
are  incompatible with very basic observations. On the other hand, so far we have witnessed a plethora of
 remarkable structures appearing in chiral 
higher spin theories as well as their striking reminiscence of their lower spin counterparts. For these reasons, we believe,
it is too premature to abandon the quest for parity invariant higher spin theories, at least until the nature and consequences of non-localities
that they feature are well understood.

With locality relaxed, the problem of the perturbative construction of higher spin interactions becomes ill-defined, see \cite{Barnich:1993vg}\footnote{A similar
statement can be proven for the light-cone deformation procedure \cite{Ponomarev:2016cwi}.}.
So, to proceed we need to find another guiding principle that will substitute locality. 
 It is natural to expect that 
 higher spin symmetries will play an important role in constructing parity invariant higher spin 
theories. In this regard, we would like to emphasise that, though, there is  a connection 
between symmetries of parity invariant and self-dual theories, it does not seem to allow to recover the former from the latter.
To illustrate it, let us consider an example of gravity. While the on-shell symmetry associated with gravity
is given by the Poincare algebra, the compete gauge-like hidden symmetry algebra for self-dual gravity is given
by the loop extension of the algebra of area-preserving diffeomorphisms\footnote{Moreover, self-dual gravity possesses
another, space-like, infinite symmetry algebra  associated with Lorentz rotations \cite{Popov:1996uu}, see also \cite{Popov:1995qb}
for the  self-dual Yang-Mills case. Similar symmetry extensions are expected for all chiral theories, but this question goes beyond the
scope of the present paper.}.
Nevertheless, it is straightforward to see that simply by constraining the generators of the algebras that we found here to be polynomials of
 appropriate degrees in $(x^-, x)$ we can reconstruct the chiral part of the Lorentz-like subalgebras of higher spin algebras
 found in AdS \cite{Boulanger:2013zza}. The problem, which is typically encountered with the remaining generators 
 \cite{Bekaert:2008sa,Sleight:2016xqq} still remains to be solved.

\section*{Acknowledgements}
 I am grateful to T. Adamo, K. Mkrtchyan,   E. Skvortsov and A. Tseytlin for fruitful discussions on different aspects of this work.
 I would also like to thank K. Mkrtchyan, E.~Skvortsov and A. Tseytlin for comments on the draft. It is also a pleasure to thank N.~Boulanger and D. Francia 
  for interesting remarks. This work was supported by the ERC Advanced grant No.290456.

\appendix

\section{Conventions}
\label{conventions}
We use the mostly plus convention for the 4d Minkowski metric
\begin{equation}
\label{4sep15}
ds^2=-(dx^0)^2+(dx^1)^2+(dx^2)^2 + (dx^3)^2.
\end{equation}
Introducing the light-cone coordinates 
\begin{align}
\notag
x^+ &\,= \frac{1}{\sqrt{2}}(x^3+x^0), & x^-&\, = \frac{1}{\sqrt{2}}(x^3-x^0),\\
\label{4sep16}
x &\,=\frac{1}{\sqrt{2}}(x^1-ix^2), & \bar x &\,= \frac{1}{\sqrt{2}}(x^1+ix^2),
\end{align}
we have 
\begin{equation}
ds^2 = 2dx^+ dx^- + 2 dx d\bar x.
\end{equation}

Our spinor-helicity conventions are chosen to be consistent with \cite{Elvang:2013cua}. Employing  the Pauli matrices
\begin{equation}
\sigma^0 = 
\left(\begin{array}{cccc}
1 &&& 0\\
0 && &1
\end{array}\right), \quad \sigma^1 = 
\left(\begin{array}{cccc}
0 & && 1\\
1& && 0
\end{array}\right), 
\quad 
 \sigma^2 = 
\left(\begin{array}{ccc}
0 && -i\\
i&& 0
\end{array}\right), \quad
 \sigma^3 = 
\left(\begin{array}{ccc}
1 && 0\\
0&& -1
\end{array}\right).
\end{equation}
 we can define momentum bi-spinors 
\begin{equation}
\label{4sep17}
q_{a\dot  b} \equiv q_\mu (\sigma^\mu)_{ a \dot  b}
=\sqrt{2}\left(
\begin{array}{ccc}
q^- && \bar q\\
 q & &-q^+
\end{array}\right)
\approx
\sqrt{2} 
\left(\begin{array}{ccc}
-\frac{q\bar q}{\beta} && \bar q\\
 q && -\beta
\end{array}\right)= - | q]_a \langle q|_{\dot b},
\end{equation}
where
\begin{equation}
\label{4sep18}
| q]_a \equiv
\frac{2^{\frac{1}{4}}}{\sqrt{\beta}}\left(\begin{array}{c}
\bar q\\
-\beta
\end{array}\right),
\qquad
\langle q |_{\dot b} \equiv \frac{2^{\frac{1}{4}}}{\sqrt{\beta}}
\left(\begin{array}{ccc}
 q&& -\beta
\end{array}\right).
\end{equation}
Then one can define spinor products
\begin{equation}
\label{4sep19}
[pq] = [ p |^a |q]_a = \varepsilon^{ab} |q]_a |p]_b, \qquad \langle p q\rangle = \langle p|_{\dot a} |q\rangle^{\dot a} = \varepsilon^{\dot a\dot b} \langle p|_{\dot a}\langle q|_{\dot b},
\end{equation}
where
\begin{equation}
\varepsilon^{ab} = \varepsilon^{\dot a\dot b} = 
\left(
\begin{array}{ccc}
0 && 1\\
-1 && 0
\end{array}
\right) = - \varepsilon_{ab}= -\varepsilon_{\dot a\dot b}.
\end{equation}
We also use the standard notation $[ij]\equiv [q_iq_j]$ and $\langle ij \rangle \equiv \langle q_i q_j \rangle$.
Rewriting spinor contractions as matrix products we find
\begin{align}
[ij]&\, = 
\frac{2^{\frac{1}{4}}}{\sqrt{\beta_j}}
\left(\begin{array}{ccc}
 \bar q_j&& -\beta_j
\end{array}\right)
\left(
\begin{array}{ccc}
0 && 1\\
-1 && 0
\end{array}
\right)
\frac{2^{\frac{1}{4}}}{\sqrt{\beta_i}}\left(\begin{array}{c}
\bar q_i\\
-\beta_i
\end{array}\right)=\frac{\sqrt{2}}{\sqrt{\beta_i\beta_j}}\bar{\mathbb{P}}_{ij},
\label{2aug1}
\\
\langle ij\rangle &\,= 
\frac{2^{\frac{1}{4}}}{\sqrt{\beta_i}}
\left(\begin{array}{ccc}
  q_i&& -\beta_i
\end{array}\right)
\left(
\begin{array}{ccc}
0 && 1\\
-1 && 0
\end{array}
\right)
\frac{2^{\frac{1}{4}}}{\sqrt{\beta_j}}\left(\begin{array}{c}
 q_j\\
-\beta_j
\end{array}\right)=-\frac{\sqrt{2}}{\sqrt{\beta_i\beta_j}}{\mathbb{P}}_{ij}.
\end{align}

This allows to establish the connection with representation (\ref{2aug3}) for cubic vertices used throughout the paper.
In particular, we have
\begin{align}
\frac{\bar{\mathbb{P}}^{\lambda_1+\lambda_2+\lambda_3}}{\beta_1^{\lambda_1}\beta_2^{\lambda_2}\beta_3^{\lambda_3}}
 = \frac{[12]^{d_{12}}[23]^{d_{23}}[31]^{d_{31}}}{\sqrt{2}^{\lambda_1+\lambda_2+\lambda_3}},
\label{1aug26}
\end{align}
where
\begin{equation}
\label{1aug27}
d_{12}\equiv\lambda_1+\lambda_2-\lambda_3,\quad d_{23}\equiv\lambda_2+\lambda_3-\lambda_1, \quad
d_{31}\equiv\lambda_3+\lambda_1-\lambda_2.
\end{equation}

Spinors satisfy various identities. The following identity, called  the Schouten identity,
\begin{equation}
\label{27aug7}
[ij][kl]=[ik][jl]+[il][kj]
\end{equation}
was used in the text.

\section{Generalised squaring procedure}
\label{gsp}
In this Appendix we propose the generalised squaring procedure that expresses off-shell chiral higher spin amplitudes in terms of
amplitudes of the self-dual Yang-Mills theory. 

At cubic level
any antiholomorphic vertex can be expressed in terms of the kinematic part of the self-dual Yang-Mills vertex as follows
\begin{equation}
\label{30aug6}
h_3^{\lambda_1,\lambda_2,\lambda_3} = \left(h_3^{1,1,-1}\right)^{\frac{\lambda_1+\lambda_2}{2}}
\left(h_3^{1,-1,1}\right)^{\frac{\lambda_1+\lambda_3}{2}}
\left(h_3^{-1,1,1}\right)^{\frac{\lambda_2+\lambda_3}{2}}.
\end{equation}
Similarly, out of kinematic numerators of higher point self-dual Yang-Mills cubic diagrams of some type $j$ one can produce 
kinematic numerators of a type-$j$ cubic diagram with arbitrary helicities on external lines as
\begin{equation}
\label{31aug7}
n_{j}^{\lambda_1,\lambda_2,\dots, \lambda_n} = \left(n_{j}^{1,1,\dots,-1}\right)^{\frac{\sum \lambda_i - (n-2)\lambda_n}{2(n-2)}}
\dots 
\left(n_{j}^{-1,1,\dots,1}\right)^{\frac{\sum \lambda_i - (n-2)\lambda_1}{2(n-2)}}.
\end{equation}
Here we additionally indicated helicities on external lines  associated  with each numerator. 

As it is not hard to see,
representation (\ref{31aug7})
also fixes  all helicities on internal lines. More precisely, if we consider an internal line that breaks the diagram into
subdiagrams ${\cal A}_{L}$
and ${\cal A}_{R}$, then representation (\ref{31aug7}) implies that helicity $\lambda_{LR}$ that flows through the propagator from ${\cal A}_R$
to ${\cal A}_L$ is  
\begin{equation}
\label{31aug8}
\lambda_{LR} = \frac{N_L-1}{N_L+N_R-2}\sum_R \lambda - \frac{N_R-1}{N_L+N_R-2}\sum_L \lambda.
\end{equation}
Here $N_L$ and $N_R$ are the numbers of external lines in subdiagrams ${\cal A}_L$ and ${\cal A}_R$
and the sums give total helicities ingoing these lines. 

This condition turns out to be satisfied for cubic diagrams of the coloured and the Poisson chiral higher spin theories.
Indeed, for the coloured chiral higher spin theory the constraint $\lambda_1+\lambda_2+\lambda_3 =1 $ at each vertex 
implies 
\begin{equation}
\label{31aug9}
\sum_L \lambda= N_L-1-\lambda_{LR}, \qquad \sum_R\lambda = N_R-1 +\lambda_{LR}, 
\end{equation}
which, in turn, leads to (\ref{31aug8}).
Similarly, for the Poisson chiral higher spin theory one has the constraint $\lambda_1+\lambda_2+\lambda_3 =2 $
at each  vertex entailing 
\begin{equation}
\label{31aug10}
\sum_L \lambda= 2(N_L-1)-\lambda_{LR}, \qquad \sum_R\lambda = 2(N_R-1) +\lambda_{LR}, 
\end{equation}
which is still consistent with (\ref{31aug8}).

In other words, we conclude that by using the generalised squaring procedure (\ref{31aug7}) one can generate
all amplitudes of the coloured (\ref{28aug2}) and the Poisson chiral higher spin theories (\ref{29aug4}) from self-dual 
Yang-Mills amplitudes,
\begin{equation}
\label{31aug11}
{\cal A}_n = g^{n-2}\sum_{j} \frac{n^{\lambda_1,\dots,\lambda_n}_j}{D_j}.
\end{equation}
At the same time, to produce amplitudes of the chiral higher spin theory (\ref{27aug2}),
the amplitudes of the self-dual Yang-Mills are not sufficient. One natural way out is to add to the theory the Born-Infeld vertex.
This generalisation is straightforward, but more cumbersome and will not be considered here.

\section{On local obstruction to higher spin interactions in flat space}
\label{locobstr}
Here we briefly review the results found in \cite{PonSkv:2016} in a joint work with E. Skvortsov.

As it was already mentioned in the body of the text, there is a consistent chiral theory that involves only 
a minimal coupling of spin $\lambda \ge 0$ field to self-dual gravity, that is only 
vertices $h_3^{\lambda,-\lambda,2}$, $h_3^{2,-2,2}$ with couplings $C^{\lambda,-\lambda,2}=C^{2,-2,2}=\ell$  are
non-vanishing. 
 In \cite{PonSkv:2016} it was shown that  this theory does not have a local parity invariant completion.
 Note, that contrary to the standard $S$-matrix no-go arguments, which lead to the conclusion that the $S$-matrix 
 in the theory is trivial, 
 in \cite{PonSkv:2016} it was found that a local theory with a parity invariant gravitational coupling of higher spin fields
  does not exist irrespectively of the form of the $S$-matrix.
  As it will be clear from the discussion
 below, a quartic vertex, that can potentially compensate for a failure of the cubic action to be Lorentz invariant, has 
 fixed homogeneity degrees in momenta $q$ and $\bar q$. In this case locality is defined simply as a requirement that 
 the quartic vertex is a polynomial of the respective degrees in these components of momenta.

To start, for a parity invariant completion we need to add non-vanishing cubic holomorphic vertices
$h_3^{\lambda,-\lambda,-2}$ and $h_3^{2,-2,-2}$ with the coupling constants  $\tilde C^{\lambda,-\lambda,2}=\tilde C^{2,-2,2}=\ell$ 
and check whether consistency conditions (\ref{2aug7}), (\ref{2aug9}) that we copy here for
convenience 
\begin{equation}
\label{4sep20}
[H_3,J_3]+[H_4,J_2]=0,
\end{equation}
\begin{equation}
\label{4sep21}
[H_3,\bar J_3]+[H_4,\bar J_2]=0 
\end{equation}
can be satisfied for any $h_4$. 

Let us focus on  (\ref{4sep20}) and a contribution to $[H_3,J_3]$ involving fields 
\begin{equation}
\label{4sep23}
\Phi^\lambda(q_1)\Phi^\lambda(q_2)\Phi^{-\lambda}(q_3)\Phi^{-\lambda}(q_4).
\end{equation}
This contribution is non-vanishing because 
$C^{\lambda,-\lambda,2}\ne 0$ and $ \tilde C^{\lambda, -\lambda,-2}\ne 0$ and  is linear in $\bar q$
and bilinear in $q$. To be more general, we  consider a possibility that vertices 
$h_3^{\lambda,\lambda,-2\lambda+2}$,  $h_3^{-\lambda,-\lambda,2\lambda-2}$, 
$h_3^{\lambda,\lambda,-2\lambda-2}$ and $h_3^{-\lambda,-\lambda,2\lambda+2}$ are
also non-zero as they produce similar contributions and, in principle, may be important for consistency
of the whole theory. Assuming that the respective couplings give the parity invariant theory, but otherwise are
 arbitrary, we obtain two additional parameters to tune.

On the other hand, $h_4$ that can compensate this contribution in (\ref{4sep20}) is also quite constrained.
First, it is not hard to see that it should be linear in $q$ and $\bar q$. Next, we take into account that 
$q$ and $\bar q$ can enter the Hamiltonian only within combinations $\mathbb{P}_{ij}$ and $\bar{\mathbb{P}}_{ij}$,
see (\ref{1aug17}). Using momentum conservation one can express all  $\mathbb{P}_{ij}$ as linear 
combinations of $\mathbb{P}_{12}$ and $\mathbb{P}_{34}$ with $\beta$-dependent factors. Explicit formulas
can be found, for example, in \cite{Bengtsson:2016jfk}. The analogous result holds for $\bar{\mathbb{P}}_{ij}$.
Eventually, this implies that   $h_4$ can be written as a linear combination of 
\begin{equation}
\label{4sep22}
\mathbb{P}_{12}\bar{\mathbb{P}}_{12}, \qquad \mathbb{P}_{12}\bar{\mathbb{P}}_{34}, \qquad \mathbb{P}_{34}\bar{\mathbb{P}}_{12}, \qquad \mathbb{P}_{34}\bar{\mathbb{P}}_{34}
\end{equation}
with $\beta$-dependent prefactors. This $\beta$ dependence has a fixed homogeneity degree, which follows from the 
last equation in (\ref{1aug16}). Moreover, $\beta$ themselves satisfy the momentum conservation condition. 
These two constraints together leave only two independent variables out of four $\beta$'s.

To summarise, $h_4$ that can, in principle, compensate the $[H_3,J_3]$ contribution produced by the minimal gravitational
interactions of helicities $\lambda$, $-\lambda$ to the sector (\ref{4sep23}) is fixed up to four functions of two 
variables. Consistency condition (\ref{4sep20}) gives an overdetirmined system of linear first order differential equations for these functions
with an inhomogeneous part produced by $[H_3,J_3]$. By verifying the associated integrability conditions for $[H_3,J_3]$ one finds 
that for $\lambda>2$ this system has no solutions even if we add vertices $h_3^{\lambda,\lambda,-2\lambda+2}$,  $h_3^{-\lambda,-\lambda,2\lambda-2}$, 
$h_3^{\lambda,\lambda,-2\lambda-2}$ and $h_3^{-\lambda,-\lambda,2\lambda+2}$. In turn, for gravity and Yang-Mills theory this test gives a
positive result, so, not surprisingly, these theories exist in light-cone gauge. 
Note that recently in a similar manner  Yang-Mills theory was derived purely from the light-cone deformation procedure \cite{Ananth:2017pio}.

Finally, let us mention a related no-go result. The coloured chiral higher spin theory turns out to be not only Poincare invariant, but also 
conformally invariant. In \cite{Metsaev:1991thesis} it was shown that this theory does not have a local, conformal and parity invariant completion.

\bibliography{KHSA}
\bibliographystyle{utphys}
\end{document}